\documentclass[a4paper,superscriptaddress,twocolumn,showpacs,amsmath,amssymb,floatfix,longbibliography]{revtex4-2}

\usepackage{graphicx}
\usepackage{amssymb}
\usepackage{float}
\usepackage{hyperref}
\hypersetup{
	bookmarks=true,         % show bookmarks bar?
	unicode=false,          % non-Latin characters in Acrobat’s bookmarks
	pdftoolbar=true,        % show Acrobat’s toolbar?
	pdfmenubar=true,        % show Acrobat’s menu?
	pdffitwindow=false,     % window fit to page when opened
	pdfstartview={FitH},    % fits the width of the page to the window
	pdftitle={My title},    % title
	pdfauthor={Author},     % author
	pdfsubject={Subject},   % subject of the document
	pdfcreator={Creator},   % creator of the document
	pdfproducer={Producer}, % producer of the document
	pdfkeywords={keyword1, key2, key3}, % list of keywords
	pdfnewwindow=true,      % links in new PDF window
	colorlinks=true,       % false: boxed links; true: colored links
	linkcolor=blue,          % color of internal links (change box color with linkbordercolor)
	citecolor=blue,        % color of links to bibliography
	filecolor=blue,         % color of file links
	urlcolor=blue        % color of external links
}
\usepackage{xcolor}
\usepackage{caption}
\usepackage{booktabs}
\usepackage{multirow}
\usepackage{tabularx}
\newcolumntype{C}{>{\centering\arraybackslash}X} % centered text in C columns
\newcolumntype{L}{>{\raggedright\arraybackslash}X} % centered text in L columns

\newcommand{\scur}{\mbox{\scriptsize\textit{\textcent}}}
\newcommand{\cur}{\mbox{\textit{\textcent}}}

\begin{document}

\title{Opinion formation in the world trade network}

\author{C\'elestin Coquid\'e}
\affiliation{\mbox{\'Equipe de Physique Th\'eorique,
    Institut UTINAM, Universit\'e de Franche-Comt\'e,
    CNRS, Besançon, France}}
    \affiliation{\mbox{Équipe Data Mining and Machine Learning, LIRIS, Université Lyon 1, CNRS / INSA Lyon, Villeurbanne, France}}
\author{Jos\'e Lages}
\affiliation{\mbox{\'Equipe de Physique Th\'eorique,
    Institut UTINAM, Universit\'e de Franche-Comt\'e,
    CNRS, Besançon, France}}
\author{Dima L. Shepelyansky}
%\homepage[]{http://www.quantware.ups-tlse.fr}
\affiliation{\mbox{Laboratoire de Physique Th\'eorique, 
Universit\'e de Toulouse, CNRS, UPS, 31062 Toulouse, France}}

%\date{today}
\date{December, 2023}

\begin{abstract}
We extend the opinion formation approach to probe the world influence of economical organizations. Our opinion formation model mimics a battle between currencies within the international trade network. Based on the United Nations Comtrade database, we construct the world trade network for the years of the last decade from 2010 to 2020. We consider different core groups constituted by countries preferring to trade in a specific currency. We will consider principally two core groups, namely, 5 Anglo-Saxon countries which prefer to trade in US dollar and the 11 BRICS+ which prefer to trade in a hypothetical currency, hereafter called BRI, pegged to their economies.
	We determine the trade currency preference of the other countries via a Monte Carlo process depending on the direct transactions between the countries. The results obtained in the frame of this mathematical model show that starting from year 2014 the majority of the world countries would have preferred to trade in BRI than USD. The Monte Carlo process reaches a steady state with 3 distinct groups: two groups of countries preferring, whatever is the initial distribution of the trade currency preferences, to trade, one in BRI and the other in USD, and a third group of countries swinging as a whole between USD and BRI depending on the initial distribution of the trade currency preferences. We also analyze the battle between 3 currencies: On one hand, we consider USD, BRI and EUR, the latter currency being pegged by the core group of 9 EU countries. We show that the countries preferring EUR are mainly the swing countries obtained in the frame of the two currencies model.
	On the other hand, we consider USD, CNY (Chinese yuan), OPE, the latter currency being pegged to the major OPEC+ economies for which we try to probe the effective economical influence within international trade. 
	Finally, we present the reduced Google matrix description of the trade relations between the Anglo-Saxon countries and the BRICS+.
	\end{abstract}

%\pacs{05.45.Mt, 67.85.Hj,  47.27.-i, 72.15.Rn}
%05.45.-a Nonlinear dynamics and chaos
%67.85.Hj 	Bose-Einstein condensates in optical potentials
%47.27.-i 	Turbulent flows
%72.15.Rn 	Localization effects (Anderson or weak localization) 
%
%47.35.-i 	Hydrodynamic waves
%47.35.Bb 	Gravity waves 
%89.75.-k 	Complex systems 

\maketitle

\section{Introduction}

The process of opinion formation is at the foundation
of functioning of democratic societies \cite{zaller92}. The development of social networks,
characterized by scale-free properties (see e.g. \cite{fortunato09,dorogovtsev10}),
makes even more important the investigation and analysis of such a process.
Various voter models had been developed for the analysis of opinion formation
as described in
\cite{fortunato09,galam86,liggett99,sznajd00,galam05,sood05,watts07,galam08,schmittmann10, biswas12, crokidakis14, biswas23}.
It is natural to assume that a given voter  opinion is influenced or even
determined by the opinions of directly linked neighbors, being similar to
the spin magnetization in the Ising model: if the spins neighboring a spin, namely a voter, are mainly up-oriented,
this spin also turns up, or if the neighboring spins are down-oriented, then the spin turns down.
Such an approach to opinion formation
on regular lattices and complex networks had been applied in many cases and
analyzed in the above cited publications. At the same time in \cite{zaller92},
it was argued that a social elite has a significant influence
on opinion formation. This feature was taken into account in \cite{kandiah12,eom15}
by attributing to each spin a weight proportional
to the PageRank probability of each node (or voter, or spin).
Such a PageRank probability is determined from the PageRank algorithm
applied to the Google matrix of a directed social network or other complex network.

The PageRank algorithm, invented in \cite{brin98}, allows to compute efficiently
the eigenvector of the Google matrix of a directed network corresponding to the leading eigenvalue. According to the Perron-Frobenius theorem,
the components of this vector are positive and give the stationary probabilities to
find a random surfer on a given node after a long series of jumps from node to node on the network.
The Google matrix construction is based on Markov chain over the network with transition probabilities determined by the number
of links from node to the other nodes (see details in \cite{brin98,langville06}).
The nodes with the highest PageRank probabilities
correspond to the most influential nodes in the network.
Such nodes were associated with social elite and it was shown, in agreement with the analysis presented in \cite{zaller92}, that
they significantly influence the opinion formation
on social and other directed networks \cite{kandiah12,eom15,frahm19}.

Various properties of real directed networks,
their Google matrices with their spectra and eigenstates
are described in \cite{ermann15a}. Among such complex networks an interesting
example is the World Trade Network (WTN) constructed from
the United Nations (UN) Comtrade database \cite{comtrade}
which gathers the annual commercial transactions between world countries.
The properties of the WTN Google matrix were studied with details in
\cite{benedictis11,ermann11,ermann15b,coquide19}. An interesting new element of the WTN
is that not only the PageRank vector plays an important role
but also the additional CheiRank vector \cite{chepelianskii10,zhirov10},
which represents the PageRank vector of the WTN for which the direction of the transactions are inverted. Indeed, the PageRank and CheiRank probability of a node are
approximately proportional to the number
of ingoing and outgoing links of this node, respectively \cite{langville06,ermann15a}.
Usually, e.g., for the World Wide Web (WWW) \cite{brin98,langville06}, the
outgoing links are not very important since they can be easily modified at the node level. However, in the WTN, ingoing and outgoing
trade transactions correspond to import or export of a given country
and definitely both play very important roles. Thus, both the PageRank and the CheiRank
vectors are essential for the analysis of WTN flows \cite{ermann11,ermann15b,coquide19}.

It is rather natural to apply the approach of opinion formation to the WTN.
For the WTN, we consider that a country can have its own opinion to perform a trade with a certain group of countries with one preferred currency (e.g., US dollar, USD)
and with other group of countries with another currency (e.g., Chinese yuan, CNY). It is possible to study the opinion evolution of the countries assuming that initially a given
fraction $f_i^{\rm USD}$ of all the countries have a trade currency preference (TCP) for USD while the remaining fraction $f_i^{\rm CNY}=1-f_i^{\rm USD}$ initially prefers to trade in CNY. Taking these initial distributions as random, a Monte Carlo iteration step
is performed {\it asynchronously}: for a given country, a new TCP is determined according to a certain linear combination of opinions of its neighbors, and this country possibly adopts
the opinion of the majority of its neighbors.
This procedure is performed sequentially for every countries picked at random and at the end of such asynchronous Monte Carlo iterations,
we obtain the final TCP distributions and consequently the final fractions
$f_f^{\rm USD}= 1-f_f^{\rm CNY}$ of countries preferring to trade in USD or CNY.
Such an approach was applied for the WTN
for years 2010 - 2020 showing that
around 2014 the major fraction of countries
would prefer to trade in CNY \cite{coquide23a}.
It is interesting to note that a similar Monte Carlo
iterations are used in the models of associative memory,
even if the linear condition for spin orientation
imposed by neighbors is
somewhat different there (see e.g. \cite{hopfield82,benedetti23}).

For the WTN, it is also possible to consider
the case of three currencies (3 possible TCPs),
e.g. USD, BRI and EUR, where BRI is a hypothetical currency pegged to the BRICS economies \cite{coquide23b}. Here, the TCP of a country is modeled as a spin which takes its orientation accordingly to the distribution of the three possible TCPs among its neighbors. 
It was also assumed \cite{coquide23b} that
for each possible currency there are core groups of countries which always have a fixed TCP: USD for 5 Anglo-Saxon countries,
the {\it hypoth\'etique} BRI currency
for the 5 BRICS and
EUR for the kernel of 9 EU countries.
The kernel of 9 EU countries was chosen
following the economic studies
reported in \cite{saint18,loye21}.
We note that the presence of the core groups with
fixed opinion reminds somehow the case
of the Sznajd opinion formation model \cite{sznajd00},
which captures the known feature of trade unions
slogan {\it united we stand, divided we fall}.

In this paper, we review the opinion formation approach
applied to the WTN with two or three currencies.
We apply this approach mainly to the case
of the currency battle between USD and BRI through the core groups of 5 Anglo-Saxon countries
and of the 11 countries which are expected to form
the new BRICS+ group from 2024 \cite{guardian23,bricswiki}.
We also consider other cases with three currencies and with other core groups for the years 2010-2020.
We also analyze the effective transactions
between the countries belonging to the core groups using the reduced Google matrix (REGOMAX) approach applied to the WTN \cite{frahm16,coquide19}.

It should be noted that from the Bretton Woods agreement in 1944
till last years, USD kept the dominant position in international trade \cite{bretton}. However,  recently, there is a clear tendency for certain countries
to perform trade in other currencies. For example,
Saudi Arabia considers to use Chinese yuan (CNY) instead of
USD for oil sales to China \cite{wallstrj}. Also CNY becomes the most traded foreign
currency at the Moscow Exchange \cite{wallstrj}. The Brazil-China authorities summit in
April 2023 pushed forward the possibility of launching a new currency BRI pegged to the 5 BRICS economies to end the trade dominance of USD
\cite{bricurrency}.
From January 2024, BRICS is expected to expand to 6 new countries
leading to the appearance of the new BRICS+ group \cite{guardian23,bricswiki} whose creation can significantly affect international trade, notably with the hypothetical appearance of its own BRI currency.
Thus the analysis of the WTN with opinion formation approach in the frame of this de-dollarization attempt era is rather interesting and timely.

The article is composed as follows:
Section II describes the data sets, the opinion formation model on the WTN
and the Google matrix construction;
Section III presents the results of the battle between two currencies, USD and BRI, which are supported by Anglo-Saxon countries and the BRICS+, respectively;
Section IV depicts the trade interactions between
these countries obtained in the frame of REGOMAX algorithm;
Section V reports the results of the three currencies battle between USD, BRI and EUR and then USD, CNY and OPE (a hypothetical petrocurrency pegged to the major OPEC+ economies);
discussion and conclusion are given in Section VI. The Appendix presents additional figures
supporting the main results presented in the article. 

\section{Model description and data sets}
%sec2

We use the UN Comtrade database \cite{comtrade} which gathers the yearly 
volumes of bilateral commercial transactions of about ten thousands of 
products.
We consider international trade between {$N=194$} countries over 
the decade $2010$-$2020$.
We use the aggregated money matrix element $M_{cc'}$ which gives the 
volume of goods, expressed in USD, exported from the country $c'$ to the 
country $c$ during a given year.
The total volume of commodities imported by and exported from the 
country $c$ is then
$M_{c}=\sum_{c'}M_{cc'}$
and
$M^{*}_{c}=\sum_{c'}M_{c'c}$, respectively. The total volume of goods 
exchanged in a given year is $M=\sum_c M_c=\sum_c M^{*}_c$.
The quantities
\begin{equation}
	S_{cc'}=\displaystyle\frac{M_{cc'}}{M^{*}_{c'}}\quad\mbox{ and }\quad
	S_{cc'}^{*}=\displaystyle\frac{M_{c'c}}{M_{c'}}\label{eq:S}
\end{equation}
give, respectively, the fraction of the total volume exported from the 
country $c'$ which is effectively imported by the country $c$ and the 
fraction of the total volume imported by the country $c'$ which is 
effectively exported from the country $c$. These quantities measure the 
relative importance of the country $c$ in the exports and the imports of 
the country $c'$. The ImportRank and the ExportRank, i.e.,
\begin{equation}
	P_{c}=\displaystyle\frac{M_{c}}{M}
	\quad\mbox{ and }\quad
	P^*_{c}=\displaystyle\frac{M^{*}_{c}}{M}
\end{equation}
quantify, respectively, the relative importance of the imports by and 
the exports from country $c$ with respect to the total volume of goods 
exchanged in the WTN in a given year.

\subsection{Asynchronous Monte Carlo determination of trade currency 
	preferences}
\label{sec:asynchron}

Let us consider an asynchronously Monte Carlo steps:

\textit{Step 0} - A TCP is randomly assigned to each country. Hence, a 
given country $c$ initially prefers to trade either with the currency $\cur_-$ or with the other $\cur_+$. An initial world distribution of the TCPs is then obtained.

\textit{Step 1} - A country $c$ with an initial TCP $\cur_0$, i.e., either $\cur_-$ or $\cur_+$, is picked at random. 
For this country, we compute the following TCP score
\begin{equation}
	Z_c = \sum_{c' \neq c} \sigma_{c'} \left(S_{c'c} + S^*_{c'c}\right) \displaystyle\frac{\left(P_{c'} +P^*_{c'}\right)}2
	\label{eq1}
\end{equation}
where the sum is performed over all the 
countries $c'$ which are economical partners of the country $c$. The $\sigma_{c'}$ parameter is equal to $-1$ if the country $c'$ TCP is $\cur_-$ and $+1$ if the country $c'$ TCP is $\cur_+$.
The country $c$ adopts then either the TCP $\cur_-$ if $Z_c<0$ or the TCP $\cur_+$ if $Z_c\geq0$. Consequently, the TCP of country $c$ possibly changes from the previous step currency 
$\cur_0$ to the new currency $\cur_1$. A second 
country is picked at random for which a new TCP-score (\ref{eq1}) is 
computed. Following the same rule as the one used for the first country, 
the second country TCP possibly changes. And so on, until a new TCP-score 
(\ref{eq1}) is computed for the last picked country, and possibly 
the TCP of this country changes. At the end of the \textit{step 1}, if 
the world distribution of the TCPs is the same as the \textit{step 0} 
one, then the Monte Carlo iteration is stopped, otherwise a \textit{step 
	2}, similar to the \textit{step 1}, is initiated based on the new TCP 
world distribution obtained at the end of the \textit{step 1}. The 
asynchronous Monte Carlo iterations stop at \textit{step n} such as the 
TCP world distribution obtained at the end of \textit{step n} is the 
same as the one obtained at the end of the \textit{step n-1}. The system 
reaches then an equilibrium characterized by a specific world distribution of 
TCPs.

\subsection{Google matrix and reduced Google matrix construction}
%sec3
\label{sec:GM}

We follow the procedure detailed in \cite{coquide19} to construct the Google matrix $G$ associated to the WTN. This Google matrix \cite{langville06} encodes the stochastic process describing a random walk on the WTN. The Google matrix element $G_{cc'}=\alpha S_{cc'}+\left(1-\alpha\right)/N$ encodes the trade interaction between country $c'$ and $c$. The probability rate to jump from the country $c'$ to the node $c$ is given by the stochastic matrix elements $S_{cc'}$ (\ref{eq:S}). Hence, the random walker follows the WTN structure with the probability $\alpha$ and can be teleported to any country with a probability $1-\alpha$. Here, the damping factor $\alpha=0.85$ ensures a unique steady state of the random walk independent on the initial conditions. The steady state is characterized by a Perron vector \cite{langville06}, $\Psi$, called the PageRank vector \cite{brin98}, defined as $G\Psi=\Psi$. The component $\psi_c$ of the PageRank vector $\Psi$ is proportional to the number of times the random walker hits the country $c$ during its journey in the WTN.

The component $\psi_c$ of the PageRank vector
$\Psi$ is proportional to the number of times the random walker forever wandering in the WTN hits the country $c$ during its journey.
Similarly to the $G$ Google matrix, it is useful to consider the $G^*$ Google matrix associated to the WTN whose all the money flows have been inverted \cite{ermann11,ermann15b,coquide19}.
The Perron vector $\Psi^*$, such as $G^*\Psi^*=\Psi^*$, is
known as the CheiRank vector of the WTN \cite{chepelianskii10,zhirov10}. While the PageRank vector $\Psi$ characterizes ingoing flows (imports), the CheiRank vector characterizes outgoing flows (exports).
%In trade outgoing and ingoing flows characterize export and import. 
Let us note that these matrix elements of the $G$ and $G^*$ matrices characterize the trade flows between the countries
related also to probability and entropy flows. 

Here, we use the reduced Google matrix (REGOMAX) analysis, borrowed from the quantum scattering theory, already used in \cite{coquide19} and described with details in \cite{frahm16}. Let us consider a set of $N_r$ selected countries. Just as the global Google matrix $G$ expresses the transactions carried out through the WTN by the $N$ world countries (with $G\Psi=\Psi$), the goal of the REGOMAX analysis is to construct a reduced Google matrix $G_{\rm R}$ associated with the $N_r$ selected countries (with $G_{\rm R}\Psi_{\rm r}=\Psi_{\rm r}$ where the $\Psi_{\rm r}$ vector gathers the components of the PageRank vector $\Psi$ corresponding to the selected countries). This $G_{\rm R}$ matrix then expresses the effective direct and indirect transactions between these countries. One of the main interests of the REGOMAX analysis is that it effectively takes into account all the information initially contained in the WTN.
According to \cite{frahm16}, the reduced Google matrix can be split into three components, i.e., $G_{\rm R}=G_{\rm rr}+G_{\rm pr}+G_{\rm qr}$. The $G_{\rm rr}$ component is the $N_{\rm r}\times N_{\rm r}$ sub-matrix of the matrix $G$ corresponding to the $N_{\rm r}$ countries. Hence, the $G_{\rm rr}$ matrix is filled with the $G_{cc'}$ components of the WTN Google matrix $G$ with countries $c$ and $c'$ belonging to the set of selected countries. While the $G_{\rm rr}$ matrix describes the direct trade links between the selected countries, the $G_{\rm pr}+G_{\rm qr}$ matrix describes the indirect trade links between the selected countries but passing through the rest of the global network made up of the other countries. The $G_{\rm pr}\sim\Psi_{\rm r}e^T$ matrix, where $e^T=\left(1,1,\dots,1\right)$, is uninteresting as it roughly contains only information about the reduced PageRank vector $\Psi_{\rm r}$. The remaining component $G_{\rm qr}$ encodes the possible hidden commercial links between the $N_{\rm r}$ selected countries.

\begin{table}[H]
\centering 
	\caption{\label{tab1}List of countries belonging to the core groups: the Anglo-Saxon group (ANGL), the BRICS+, the kernel of 9 European countries (EU9), and the OPEC+ countries (we select those producing more than 1 million oil barrels in 2022). The countries are sorted in decreasing order of $\max \left(P_c,P^{*}_{c}\right)$ where $P_{c}$ and $P^{*}_{c}$ are the ImportRank and the ExportRank of a country $c$ in 2019 WTN, respectively.}
	\begin{tabular}{|l|l|l|}
		\hline
		Core group&Country name & Currency\\
		\hline
		\hline
		\parbox[H]{2mm}{\multirow{5}{*}{\rotatebox[origin=c]{90}{Anglo-Saxon~}}}
		&United States of America & \multirow{6}{*}{USD}\\
		&United Kingdom &\\
		&Canada & \\
		&Australia & \\
		&New Zealand &\\
		\hline\hline
		\parbox[H]{2mm}{\multirow{13}{*}{\rotatebox[origin=c]{90}{BRICS+}}}
		&China & \multirow{13}{*}{BRI}\\
		&India &\\
		&Russia &\\
		&United Arab Emirates & \\
		&Brazil & \\
		&Saudi Arabia & \\
		&South Africa & \\
		&Argentina & \\
		&Egypt & \\
		&Iran & \\
		&Ethiopia & \\
		\hline
		\hline
		\parbox[H]{2mm}{\multirow{13}{*}{\rotatebox[origin=c]{90}{EU9}}}
		&China & \multirow{13}{*}{EUR}\\
		&Germany&\\      
		&France&\\
		&Netherlands&\\
		&Italy&\\
		&Belgium&\\   
		&Spain&\\  
		&Austria&\\    
		&Portugal&\\     
		&Luxembourg&\\
		\hline
		\hline
		\parbox[H]{2mm}{\multirow{13}{*}{\rotatebox[origin=c]{90}{OPEC+}}}
		&Saudi Arabia&\multirow{13}{*}{OPE}\\
		&Russia&\\
		&Iraq&\\
		&United Arab Emirates&\\
		&Kuwait&\\
		&Iran&\\
		&Mexico&\\
		&Kazakhstan&\\
		&Angola&\\
		&Nigeria&\\
		&Oman&\\
		&Algeria&\\
		&Libya&\\
		\hline
	\end{tabular}

%	\noindent{\footnotesize{\textsuperscript{1} Tables may have a footer.}}
\end{table}
\newpage
A qualitative description of the various features
of the REGOMAX algorithm and its applications
to the MetaCore network of protein-protein
interactions is given in \cite{kotelnikova22}. 

\section{Battle of two currencies: USD versus BRI}
%sec4
\label{sec:2currencies}

\begin{figure}[H]
	\begin{center}
		\includegraphics[width=\columnwidth]{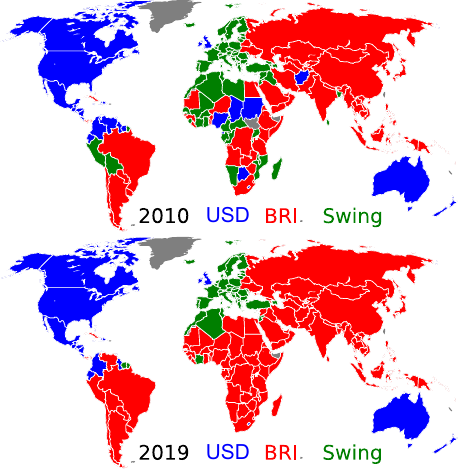}
	\end{center}
	\vglue -0.3cm
	\caption{\label{fig1}World distribution of the trade currency preferences for the years 2010 (top) and 2019 (bottom).
		%  and $G_{1}$-option of core groups of Table~\ref{tab1}.
		The countries belonging to the USD group and the BRI group are colored in blue and red, respectively. Those belonging to the swing group are colored in green.
		Countries colored in grey have no trade data reported in the UN Comtrade database for the considered year \cite{comtrade}.
		The world distribution of trade currency preferences for 2012, 2014, 2016, 2018 and 2020 are presented in Fig.~\ref{figA1}.
		%  For $G_{2}$ see Fig.~\ref{figA3}. 
	}
\end{figure}

In this Section, we discuss the results obtained for the currency battle
between a set of 5 countries of the Anglo-Saxon world (named hereafter ANGL) and the set of the 11 BRICS+ countries. The list of these countries is given in Table~\ref{tab1}.
We use the WTN built from the UN Comtrade database \cite{comtrade} for the years 2010 - 2020.
Of course, the BRICS+ was not yet formed at that epoch
but we model it as a {\it hypoth\'etique} group to characterize its
potential influence nowadays and in the future.
We assume that the countries of each of these 2 core groups, ANGL and BRICS+, always perform trade in USD and BRI, respectively.
Thus, their corresponding Ising spins $\sigma_c$ are always oriented down, i.e., $\sigma_c = -1$, for USD
and up, i.e., $\sigma_c=+1$, for BRI. Then, we consider an initial configuration of
random up and down spin orientations for the other 178 countries (194 countries minus the 16 countries listed in Tab.~\ref{tab1}) mimicking the world distribution of the initial TCPs (\textit{step 0} of the asynchronous Monte Carlo procedure described in Section~\ref{sec:asynchron}). The initial 
fraction of spins oriented down is $f_i^{\rm USD}$ and the initial fraction of spins oriented up is $f_i^{\rm BRI}=1-f_i^{\rm USD}$. As the TCPs of the 16 countries belonging to the core groups (ANGL and BRICS+) are fixed, we perform asynchronous Monte Carlo iterations on the remaining 177 countries (see section \ref{sec:asynchron}). After each step the TCP-score world distribution (\ref{eq1}) is computed until the convergence to a steady-state configuration of the TCPs over the world countries is achieved. For a given initial fraction $f_i^{\rm USD}$ of countries with a TCP for USD, we performed the asynchronous Monte Carlo iterations from $N_{\rm conf}=10^4$ different initial random spin configurations. 
The final 
fraction of spins oriented down, i.e., the final fraction of countries preferring USD, is $f_f^{\rm USD}$ and the final fraction of spins oriented up, i.e., the final fraction of countries preferring BRI, is $f_f^{\rm BRI}=1-f_f^{\rm USD}$.
The convergence to a steady-state orientation of all the spins takes place 
after $\tau$ steps (one step corresponds to the 177 asynchronous iterations and computations of (\ref{eq1}) for the 177 spins).

\begin{figure}[H]
	\begin{center}
		\includegraphics[width=\columnwidth]{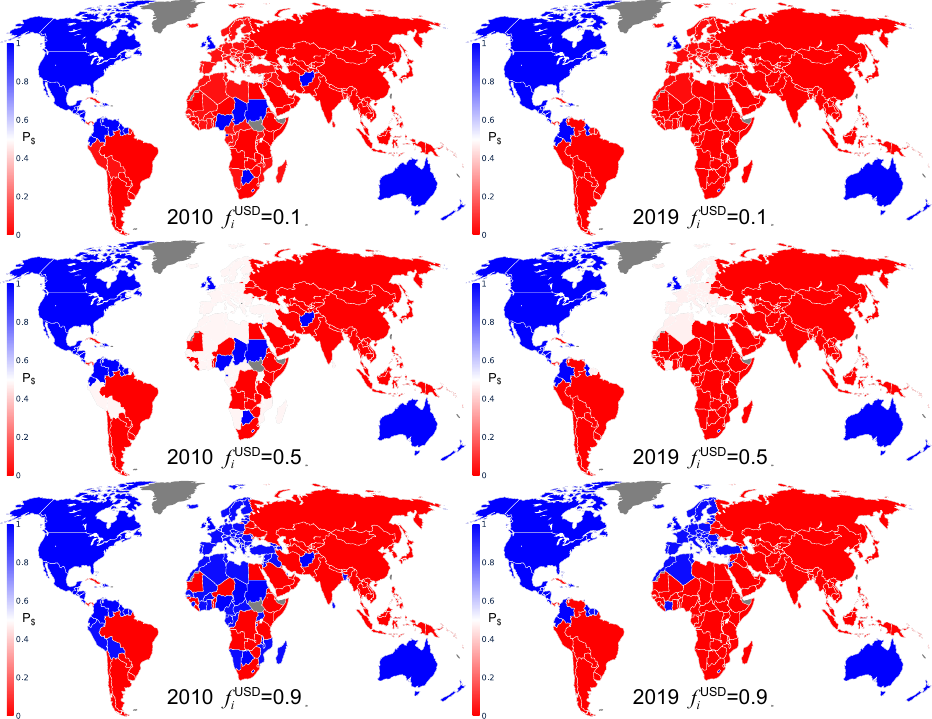}
	\end{center}
	\vglue -0.3cm
	\caption{\label{fig2}World distribution of the probability $P_\$$ that a country choose USD as its trade currency for 2010 (left) and 2019 (right), and for $f_{i}^{\rm USD}=0.1$ (top), $0.5$ (center) and $0.9$ (bottom). The colors range from red for countries which always have a TCP for BRI ($P_\$=0$) to blue for countries which always have a TCP for USD ($P_\$=1$).
		Countries colored in grey have no trade data reported in the UN Comtrade database for the considered year \cite{comtrade}.
		%Fig.~\ref{figA4}. ??
	}
\end{figure}

The main obtained results are presented in Figs.~\ref{fig1},~\ref{fig2},~\ref{fig3}
with related Figs.~\ref{figA1} and Fig.~\ref{figA2} in the Appendix. A rapid convergence of the asynchronous
Monte Carlo iterations (see section~\ref{sec:asynchron}) is shown in Fig.~\ref{figA2}. Indeed, the convergence is reached after approximately $\tau = 4$ steps (Fig.~\ref{figA2}).
In fact, only two stable values of the final fraction $f_f^{\rm USD}$ exist: $f_f^{\rm USD}=0.21$ and $0.61$ in 2010; and $f_f^{\rm USD}=0.18$ and $0.45$ in 2019 (see Fig.~\ref{fig3}).
Although these two stable values are independent of the initial fraction $f_i^{\rm USD}$, the
probability $\rho_{f_f^{\rm USD}}(f_i)$ to reach a given stable fraction value $f_f^{\rm USD}$
changes with the initial fraction $f_i^{\rm USD}$ as it is shown in Fig.~\ref{fig3}.
The two final possible outcomes for $f_f^{\rm USD}$ evolve from year to year.
The existence of two final attractors $f_f^{\rm USD}$ for the asynchronous Monte Carlo iterations implies that certain countries,
outside of the ANGL and BRICS+ core groups, always keep their TCP independent of the initial fraction $f_i^{\rm USD}$.
At the same time, there are countries, hereafter named swing countries,
that change their TCP depending on the $f_i^{\rm USD}$ value.
Thus, we have 3 groups being the USD group gathering the countries of the ANGL group and the countries with USD as final TCP, the BRI group gathering the BRICS+ and the countries with BRI as final TCP, and the swing group.
The world distribution of these groups is shown in Fig.~\ref{fig1}
for years 2010 and 2019 (the world maps for other years in the range 2010-2020
are shown in Appendix Fig.~\ref{figA1}).
There is a striking evolution from 2010 to 2019:
the number of swing countries is significantly reduces and
almost all Latin American, African and Asian countries have a TCP for BRI in 2019.
In 2019, the swing countries only comprise the EU countries, Norway, Former Yugoslavia countries, Albania, Moldavia, Azerbaijan, Turkey, Liban, Tunisia, Algeria, Morocco, Ivory Coast, and Suriname. According to the Appendix Fig.~\ref{figA1}
such a transition takes place around the year 2016.

For the swing countries, the probability to have one or the other TCP depends of the initial fraction $f_i^{\rm USD}$ as it is shown in Fig.~\ref{fig2}.
For $f_i^{\rm USD}=0.9$, all the swing countries in 2010 and 2019
prefer to trade with USD with probability close to unity
(but the number of countries with a TCP for USD is reduced from 2010 to 2019).
However, for $f_i^{\rm USD}=0.5$, in 2010 and 2019, all the European countries have a probability about $P_\$\simeq0.5$ to choose USD as trade currency.
For $f_i^{\rm USD}=0.1$, practically all the countries
would prefer to trade with BRI with a probability close to unity ($P_\$\simeq0$) on 2019
(besides countries belonging to the ANGL group, only the Central American countries, Columbia, and Ecuador would keep a TCP for USD).
\begin{figure}[H]
	\begin{center}
		\includegraphics[width=\columnwidth]{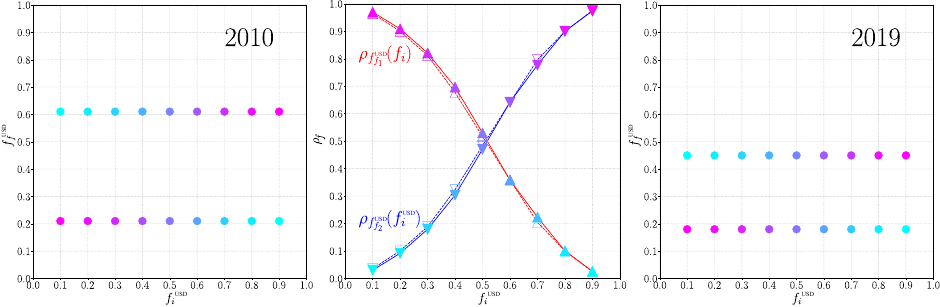}
	\end{center}
	\vglue -0.3cm
	\caption{\label{fig3}Final fraction $f_{f}^{\rm USD}$ of countries with a trade currency preference for USD versus the initial fraction $f_{i}^{\rm USD}$ of these countries for years 2010 (left panel) and 2019 (right panel). There are two possible final fractions for each considered year: $f_{f_{1}}^{\rm USD} = 0.21$  and $f_{f_{2}}^{\rm USD} = 0.61$ in 2010, and $f_{f_{1}}^{\rm USD} = 0.18$ and $f_{f_{2}}^{\rm USD} = 0.45$ in 2019. The color of the points represents the ratio of Monte Carlo process with the corresponding final state $\rho_{f_{f}}(f_{i})$, low ratio in cold blue and high ration in violet. The central panel shows evolution of $\rho_{f_{f}}(f_{i})$ with $f_{i}$. The red (blue) curve and the up (down) triangles denote the minimal (maximal) final state. The full (empty) symbols correspond to the year 2019 (2010).
	}
\end{figure}

\begin{figure}[H]
	\begin{center}
		\includegraphics[width=\columnwidth]{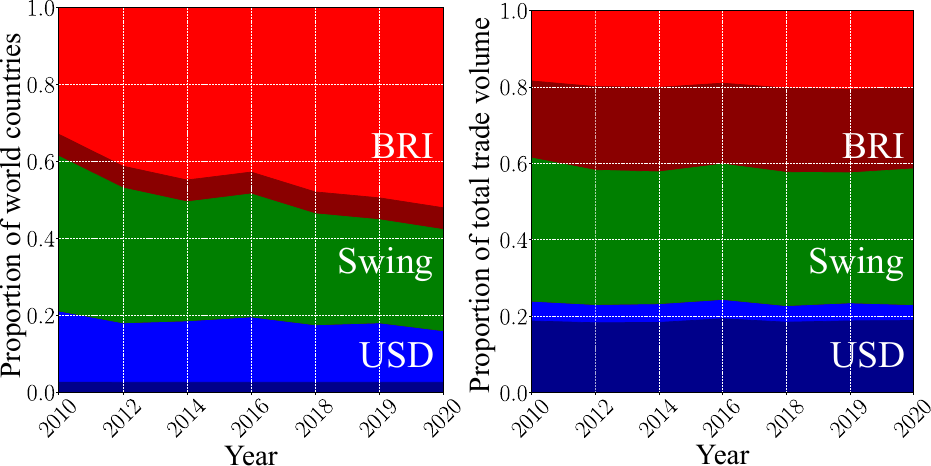}
	\end{center}
	\vglue -0.3cm
	\caption{\label{fig4}Time evolution of the size of the trade currency preference groups.
		The width of a given band corresponds to the corresponding fraction of world countries in a TCP group (left panel) and to the corresponding fraction of the total trade volume generated by this group (right panel).
		The USD group is colored in blue, the BRI group in red, and the swing group in green. Within the BRI (USD) group, the proportion corresponding to the BRICS+ (the ANGL countries) is shown in dark red (dark blue).}
\end{figure}

These results, obtained in the frame of our mathematical model,
show that the hypothetical BRI currency would becomes dominant in present day world trade. This is corroborated by the proportion of countries belonging respectively to the USD, the BRI and the swing groups (see Fig.~\ref{fig4}). 
The BRI group grows from about 38\% of the world countries in 2010 to 58\% in 2020. As the USD group only slightly decreases during this decade, the BRI group has mainly captured countries from the swing group. However, these captured countries have a somewhat small economic weight as the relative proportions of the total trade volume corresponding to these three groups stay stable during the last decade. Hence, in 2020, the BRI group represents 41\% of the total trade volume, the swing group 36\% and the USD group the remaining 23\%. For the USD group, the major trade volume is associated to the ANGL countries. By contrast, the BRICS+ only represent about a half of the total trade volume proportion of the BRI group. Contrarily to the USD  group, the BRI currency is able to influence countries with a non negligible economic weight way beyond the perimeter of the BRICS+.

%\begin{figure}[H]
%\begin{center}
%\includegraphics[width=\columnwidth]{fig5}
%\end{center}
%\vglue -0.3cm
%\caption{\label{fig5}\red{This is the figure for ALL-ANGL case. We cannot have such a figure for $f_i=0.9$ in the standard case since the standard case gives the same ANGL, BRI+ and Swing cores whatever $f_i$ is.}
	%}
%\end{figure}

We have previously considered the ANGL group formed by 5 countries of the Anglo-Saxon world.
In order to analyze how USA is influential among the other Anglo-Saxon countries, we consider the case where the ANGL group only contains the USA, the remaining Anglo-Saxon countries being now free to choose their TCP. We keep nonetheless the BRICS+ group unchanged. The results corresponding to this change are shown in Appendix Fig.~\ref{figA3} and Fig.~\ref{figA4}.
From Fig.~\ref{figA3}, we see, both in 2010 and 2019, that among Anglo-Saxon world countries, Australia and New Zealand firmly
adopt the BRI trade preference, UK enters the swing group and only Canada
keeps USD as its TCP. The evaluation of the final TCP probability for
different initial $f_i^{\rm USD}$ values is shown in Fig.~\ref{figA4}.
The world maps are similar to those of Fig.~\ref{fig2}
with the main difference that Australia and New Zealand always have a TCP for BRI
and that UK prefers USD for $f_i^{\rm USD}=0.9$, and prefers as the other European countries  BRI for $f_i^{\rm USD}=0.5$ and $0.1$ (resulting in $f_f^{\rm USD} \simeq 0.58$ and $0.16$, respectively).
These results show that the choice of the UK's TCP is similar to those of the EU countries, as one could expect for the considered pre-Brexit period,
and that Australia and New Zealand have very strong economic ties with the BRICS+ countries.

Also, the comparison of Fig.~\ref{fig1}
with Fig.~4 in \cite{coquide23a} (an article devoted to the battle USD vs. CNY in international trade) shows that in the BRICS+
the dominant economic role is played by China
(see similarity of the distributions of the swing countries in the two figures).

Finally, we make a note about the asynchronous Monte Carlo iteration process controlled by the TCP-score (\ref{eq1}). This score computed for a given country $c$ contains not only trade flows between the country $c$ and its trade partners $c'$ given by the $S_{c'c}$ matrix elements but also the factor $(P_{c'}+P_{c'}^*)$ which is the relative
economical weight of the countries $c$ direct trade partners. More important is a country in the world trade network
the higher is its relative economical weight given by its ImportRank and ExportRank. In a sense, this last factor takes
into account the world importance of a country $c'$. This is similar to
the opinion formation approach
used in \cite{kandiah12} for which the position of the neighbors of an agent in the society elite \cite{zaller92}
is taken into account to infer the opinion of this agent. Thus, the \textit{vote} of China or USA counts in the choice of the TCP of their direct partners
much more than the \textit{vote} of least developed countries.
% such as, e.g., Sudan ??TAKE one from smallest $P_c$??.
We think that this factor correctly takes into account the trade relations between the countries. 
However, in the case one object the use of this factor in the TCP-score,
we also present, in Appendix Figs.~\ref{figA5},~\ref{figA6},~\ref{figA7},
the results when this factor is neutralized by taking $P_{c'}+{P_{c'}}^*=1$ in (\ref{eq1}). In such a case, there is a larger number of steady state fractions $f_f^{\rm USD}$ as it is shown in Appendix Fig.~\ref{figA7}.
The distribution of the swing countries is shown in Appendix Fig.~\ref{figA5} for year 2010 and 2019, and with a high fraction $f_i^{\rm USD}=0.9$ of countries initially preferring USD. 
Qualitatively, we still observe a significant increase of the BRI group from 2010 to 2019
but certain countries remain in the swing group in 2019, i.e., Southeast Asia, Central Asian countries, South Korea, and Japan in Asia, few West African countries and Madagascar in Africa,
and, Chili and Peru in South America.
The probability that a country adopt a TCP for USD at the final stage of the asynchronous Monte Carlo iterations is shown,
for $f_i^{\rm USD}=0.9$, in Appendix Fig.~\ref{figA6}. Here, many of swing countries from the Fig.~\ref{figA5} adopt a USD TCP in 2010 and 2019.
Rather naturally, the steady state features are more stable and robust
when the  weight $P_{c'}+P_{c'}^*$ of the countries belonging to the trade leader elite is taken into account in the TCP-score (\ref{eq1})
and, hence we present our analysis in the frame of this relation.

\section{REGOMAX for the Anglo-Saxon group and the BRICS+}
%sec5

\begin{figure}[t]
	\begin{center}
		\includegraphics[width=\columnwidth]{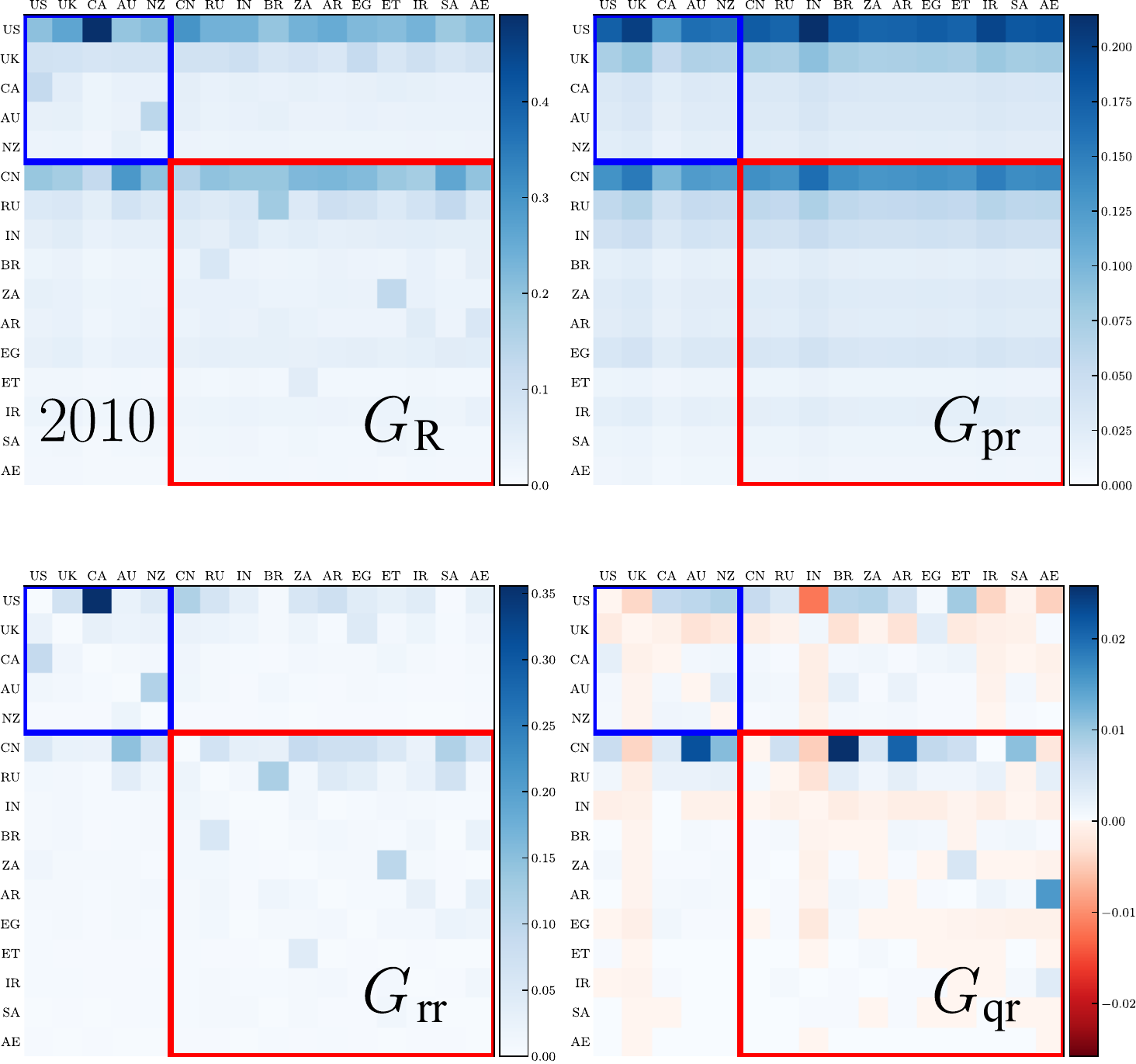}
	\end{center}
	\vglue -0.3cm
	\caption{\label{fig6}Reduced Google matrix $G_{\rm R}$ and its components
		for the ANGL countries and the BRICS+ and for the year 2010: $G_{\rm R}$ (top left), $G_{\rm pr}$ (top right),
		$G_{\rm rr}$ (bottom left) and $G_{\rm qr}$ (bottom right).
		For the $G_{\rm qr}$ matrix the relative weight of negative elements is $(W_{+} - W_{-}) / (W_{+} + W_{-}) = 0.31$,
		where $W_{+}$ and $W_{-}$ are respectively the mean of positive and negative elements (in absolute value).}
\end{figure}

\begin{figure}[t]
	\begin{center}
		\includegraphics[width=\columnwidth]{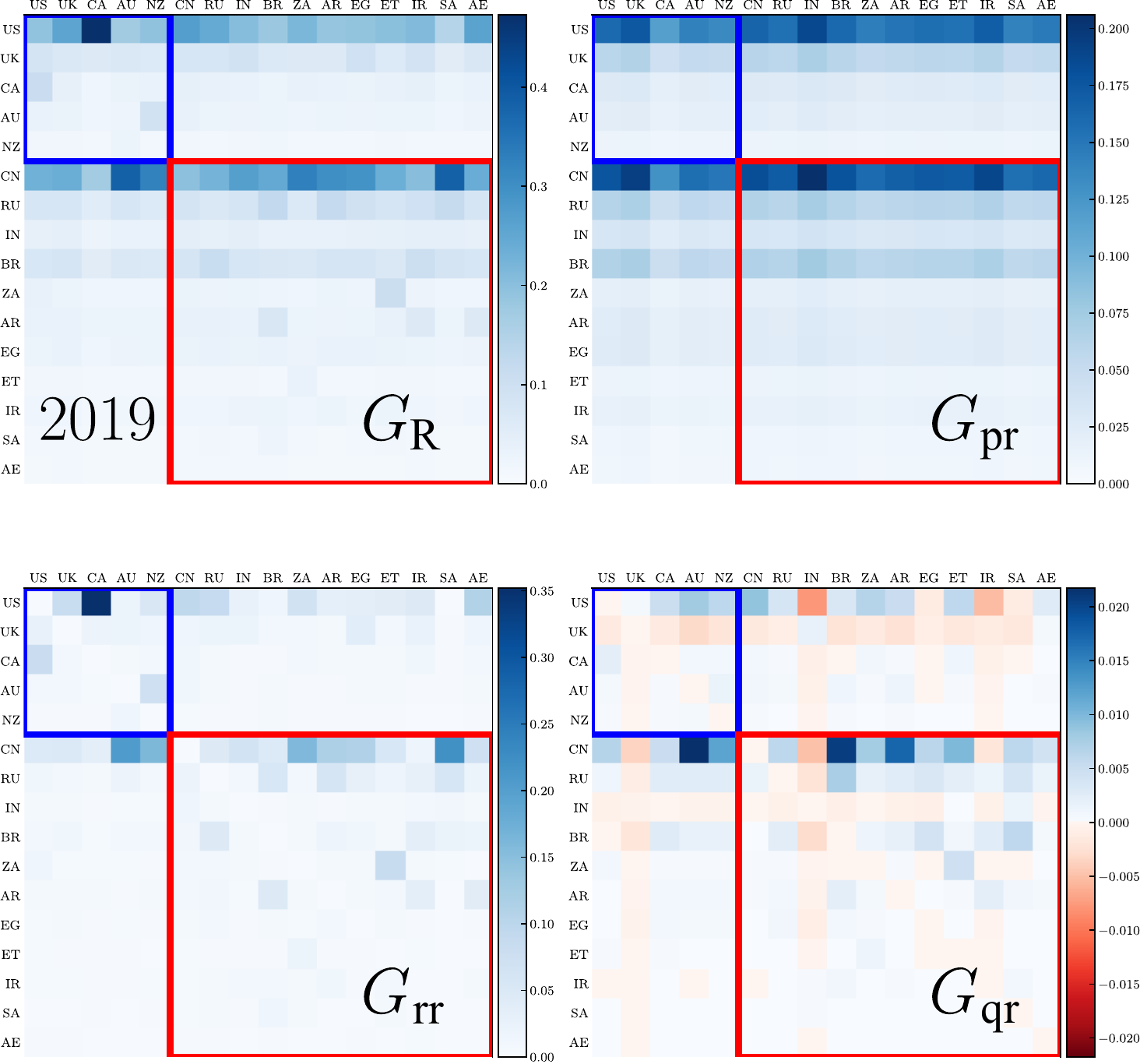}
	\end{center}
	\vglue -0.3cm
	\caption{\label{fig7}Same as in Fig.~\ref{fig6} but for the year 2019;
		here $(W_{+} - W_{-}) / (W_{+} + W_{-}) = 0.27$.}
\end{figure}

\begin{figure}[t]
	\begin{center}
		\includegraphics[width=\columnwidth]{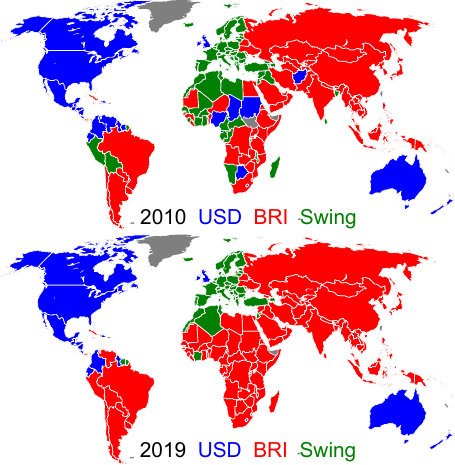}
	\end{center}
	\vglue -0.3cm
	\caption{\label{fig8}Same as Fig.~\ref{fig1}, but when the ImportRank and ExportRank probabilities in the TCP-score (\ref{eq1})
		are replaced by PageRank and CheiRank probabilities of the WTN Google matrix.}
\end{figure}

\begin{figure}[t]
	\begin{center}
		\includegraphics[width=\columnwidth]{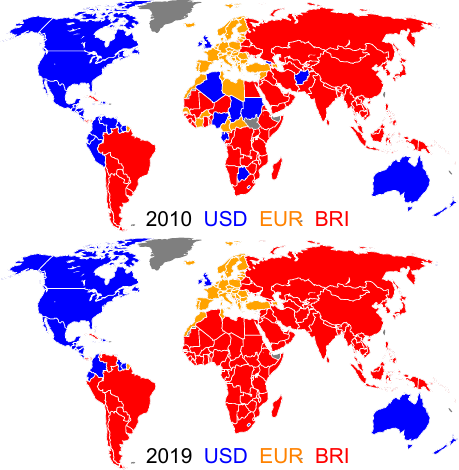}
	\end{center}
	\vglue -0.3cm
	\caption{\label{fig9}World distribution of the trade currency preferences for the years 2010 (top) and 2019 (bottom).
		%  and $G_{1}$-option of core groups of Table~\ref{tab1}.
		The countries belonging to the USD, EUR, and BRI groups are colored in blue, gold and red.
		Countries colored in grey have no trade data reported in the UN Comtrade database for the considered year \cite{comtrade}.
		%The world distribution of trade currency preferences for 2012, 2014, 2016, 2018 and 2020 are presented in Fig.~\ref{figA1}.
		%  For $G_{2}$ see Fig.~\ref{figA3}.
	}
\end{figure}

\begin{figure}[t]
	\begin{center}
		\includegraphics[width=0.5\columnwidth]{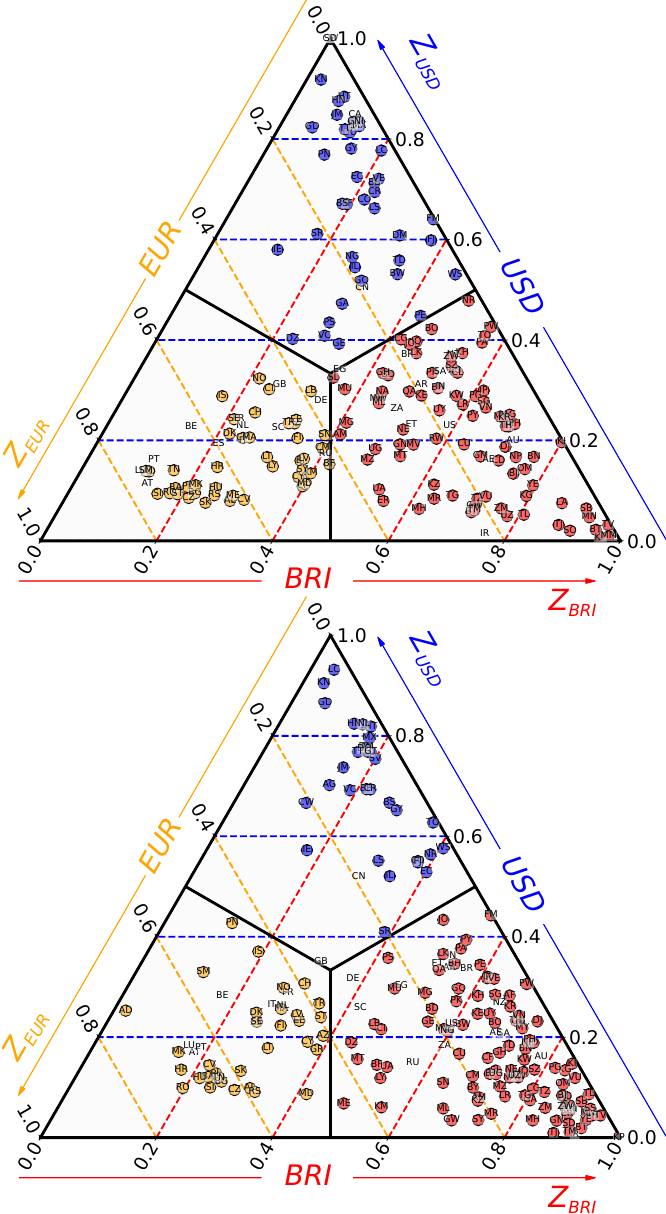}
	\end{center}
	\vglue -0.3cm
	\caption{\label{fig10}Distribution of countries' TCP scores $\left(Z_{\mbox{\tiny USD}},Z_{\mbox{\tiny EUR}},Z_{\mbox{\tiny BRI}}\right)$ for 2010 (top) and 2019 (bottom).  A country is represented by a circle. Colors are associated to TCPs, blue for USD, gold for EUR, and red for BRI. The $Z_{\mbox{\tiny USD}}$ coordinate is read along the dashed blue horizontal lines, the $Z_{\mbox{\tiny EUR}}$ coordinate along the gold dashed oblique lines, and the $Z_{\mbox{\tiny BRI}}$ coordinate along the red dashed oblique lines.
	}
\end{figure}

We present the REGOMAX description
of the trade interactions between the 5 Anglo-Saxon countries (ANGL group) and the 11 BRICS+ countries
obtained from the 2010 and 2019 WTN. The whole Google matrix $G$ of the WTN
has a size $N=194$ being the total number of countries.
As described in section \ref{sec:GM}, the REGOMAX algorithm
allows to obtain the reduced Google matrix $G_{\rm R}$ of size $N_{\rm R}=16$
which describes the trade interactions between these 16 countries
taking into account all their indirect interactions via the other
$194-16=178$ countries of the global WTN.
Thus $G_{\rm R}$ and its components characterize the integrated import flows.
The matrix $G_{\rm R}$ has three components, $G_{\rm R}= G_{\rm pr} + G_{\rm rr} + G_{\rm qr}$.
All these matrices, $G_{\rm R}$, $G_{\rm rr}$, $G_{\rm pr}$, $G_{\rm qr}$, are shown in Figs.~\ref{fig6} and \ref{fig7}
for the years 2010 and 2019, respectively.
In each core group, ANGL and BRICS+,
the countries are sorted according to their values of ImportRank $P_c$ and ExportRank $P^*_c$, namely by descending order of the $\max_c\left(P_c,P^*_c\right)$ value.
We remind that usually $G_{\rm pr}$ is close to a matrix
composed by columns filled with the PageRank vector components of the $N_{\rm R}$ countries,
$G_{\rm rr}$ encodes the direct transitions between the $N_{\rm R}$ countries
and $G_{\rm qr}$ is obtained by the sum over all the indirect pathways linking the $N_{\rm R}$ through the whole global WTN.
By definition the weight of the $G_{\rm R}$ components are
$W_{\rm R} = \sum_{ij} G_{\rm R}(i,j)/N_{\rm R}=1$ and thus $W_{\rm R}=W_{\rm rr}+W_{\rm pr}+W_{\rm qr}=1$,
where the weights of the 3 matrix components are defined
in the same way as for $G_{\rm R}$. For Figs.~\ref{fig6} and \ref{fig7},
we have $W_{\rm pr}= 0.723$ and $0.695$,  $W_{\rm rr}= 0.259$ and $0.287$,
$W_{\rm qr}= 0.018$ and $0.018 $ respectively for the years 2010 and 2019.
Thus, the main contribution to $G_{\rm R}$ comes from
$G_{\rm pr}$ and $G_{\rm rr}$, while the contribution of
$G_{qr}$, although small, provides an important
information on indirect trade transfers between the $N_{\rm R}$ countries.

From Figs.~\ref{fig6} and \ref{fig7}, we see that the strongest transition elements
form two horizontal lines for USA and China that are well visible in $G_{\rm R}$ and $G_{\rm pr}$.
This is rather natural since these two countries are at the top of PageRank probabilities
and are the most important importer in the world trade.
According to reduced Google matrix $G_{\rm R}$, the most strong trade import to USA is from Canada,
while for China it is from Australia and Saudi Arabia.
We also see that the total trade transfer from the 16 countries towards USA
($T_{\rm US} = 3.9$ in 2010 and $3.2$ in 2019) and towards
China ($T_{\rm CN} = 3.6$ in 2010 and $4.2$ in 2019) 
is significantly changed from 2010 to 2019 being enhanced for China and decreased for USA (here, $T_{c'}=\sum_c {G_{\rm R}}_{c',c}$ where $c'$ is either USA or China, and the sum is performed over the countries $c$ different from the country $c'$).

From the $G_{\rm pr}$ matrix, we see also that USA and China has important trade import 
from UK, India and surprisingly from Iran, both in 2010 and 2019. 

The contributions from the direct trade are given by the matrix $G_{\rm rr}$ components.
For USA, in 2010, the strongest imports are from Canada and from China
(but significantly smaller), and in 2019, it is still Canada and the Arab Emirates.
For China, the strongest imports are from  Australia, South Africa and Saudi Arabia in 2010 and 2019
(New Zealand becomes also important in 2019).

The indirect trade transitions are described by the matrix $G_{\rm qr}$ elements.
These matrix elements are on average smaller than those of the matrices
$G_{\rm pr}$ and $G_{\rm rr}$, however they allows to establish indirect relations between countries.
Thus, for USA, we have the most significant matrix elements
from India, Arab Emirates, Iran and UK in 2010,
and India and Iran in 2019, these elements are negative.
In principle negative elements for the $G_{\rm qr}$ matrix are
not forbidden since only the $G_{\rm R}$ elements should be positive.
Usually, negative elements are small and appear only for
$G_{\rm qr}$ (see e.g. \cite{frahm16}). Here, the negative elements
are well visible even if they still remain smaller
compared to the positive ones. Such a negativity implies that for USA
the indirect trade links between USA and India, or Iran,
reduce the direct trade transfers to USA from these countries.
For China, the strongest positive indirect links are from
Australia, Brazil and Argentina both in 2010 and 2019;
negative elements are small
(the most visible is a negative link from India in 2019).

It is possible to consider the REGOMAX results in more depth, or to apply them to other groups  of countries, but we will not opt for such
extensions since the main objective of this work is an analysis of the opinion formation in the WTN.

Finally, we note that in the TCP-score (\ref{eq1})
the probabilities can be defined not from the trade volume fractions of
export and import, i.e., from the ImportRank and the ExportRank, but from the PageRank and CheiRank probabilities
of the global Google matrix $G$ constructed from the trade of 194 countries. The countries' TCPs  
obtained with such PageRank and CheiRank probabilities implemented in the TCP-score (\ref{eq1})
for 2010 and 2019 are shown in Fig.~\ref{fig8}. They are very close to those
of Fig.~\ref{fig1} obtained with ImportRank and ExportRank probabilities.
This shows that the obtained results are very robust and
are not significantly affected by moderate modifications
of the measure of the global importance of the export and export capabilities of a country we use in the TCP-score definition (\ref{eq1}).

\section{Three currencies case}
%sec6

\begin{figure}[t]
	\begin{center}
		\includegraphics[width=\columnwidth]{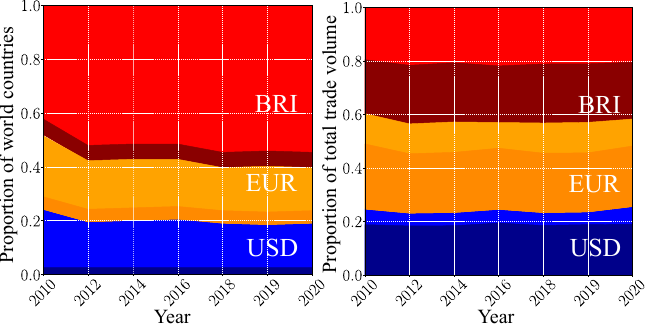}
	\end{center}
	\vglue -0.3cm
	\caption{\label{fig11}Time evolution of the size of the trade currency preference groups.
		The width of a given band corresponds to the corresponding fraction of world countries in a TCP group (left panel) and to the corresponding fraction of the total trade volume generated by this group (right panel).
		The USD group is colored in blue, the BRI group in red, and the EUR group in gold. Within the BRI (USD) [EUR] group, the proportion corresponding to the BRICS+ (the ANGL countries) [the EU9 group] is shown in dark red (dark blue) [dark gold].
	}
\end{figure}

Let us now consider three currencies, namely USD, EUR and BRI. We keep the USD and the BRI core groups, namely the ANGL group and the BRICS+ (see Table~\ref{tab1}). The EUR core group is the EU9 group constituted by Austria, Belgium, France, Germany, Italy, Luxemburg, Netherlands, Portugal, and Spain \cite{saint18}. Hence, the countries of the ANGL group keep trading in USD, the countries of the EU9 group in EUR and the BRICS+ in BRI. For each country $c$, we compute the following TCP-score for each currency $\cur\in\left\{\rm USD, EUR, BRI\right\}$ 
\begin{equation}
	Z_{c,\scur} = \frac{\sum^{(\scur)}_{c'\neq c} \left(S_{c'c}+S^{*}_{c'c}\right)\left(P_{c'}+P^{*}_{c'}\right)}{\sum_{c'\neq c} \left(S_{c'c}+S^{*}_{c'c}\right)\left(P_{c'}+P^{*}_{c'}\right)}
	\label{eq2}
\end{equation}
where the sum $\sum_{c'\neq c}^{(\scur)}$ is performed over all the countries which are commercial partners of the country $c$ and which prefer to trade with the currency $\cur$. For a given country $c$, the sum of these scores over the different considered currencies is equal to one, i.e.,
$\sum_{\scur}
Z_{c,\scur}=1$.
The country $c$ adopts then the trade currency for which the TCP-score is maximum, i.e., $\cur_1$ such as $Z_{c,\scur_1}=\max_{\scur}Z_{c,\scur}$.
The asynchronous Monte Carlo procedure (see section \ref{sec:asynchron}) is performed with the above defined TCP-score (\ref{eq2}).
For a given year among those considered (2010-2020), the final steady state distribution of the TCPs is unique and does not depend on the initial distribution of the TCPs. By contrast with the two currency model (see section~\ref{sec:2currencies}), there is no swing country whose the TCP changes according to the initial distribution of the TCPs.

The world distributions of the TCPs for the years 2010 and 2019 are shown in Fig.~\ref{fig9}. We observe that both in 2010 and 2019, the USD group gathers, beside the ANGL group countries, the Central American countries and the some countries of the Northern of the South America; the EUR group gathers, besides the EU9 group countries, the European countries delimited at the East by the NATO border; the BRI group gathers, beside the BRICS+, most of the Asian and South American countries. The main difference between the 2010 and 2019 years is the situation of the African continent. Although this continent appears fragmented in 2010, echoing the post-1989 era battle of influence between France, USA, Russia and China, in 2019,  the whole African continent appears to be under the influence of the BRI with the exception of Morocco and Tunisia. By comparison with the two currencies model (see section~\ref{sec:2currencies}), the introduction of a third currency, namely the EUR, crystallizes the swing group: undecided countries in the
two currencies model now adopt a fixed currency preference independent of the initial distribution of the TCPs. Moreover, the new EUR group mostly corresponds to the swing group (see Fig.~\ref{fig2}).

The distribution of the TCP-scores (\ref{eq2}) over the range $\left[0,1\right]\times\left[0,1\right]\times\left[0,1\right]$ is shown in Fig.~\ref{fig10}. From 2010 to 2019, we observe a clear depletion of the USD and the EUR sectors to the profit of the BRI sector. Whereas some countries are strongly influenced by USD or BRI with $Z_{\rm USD}>0.8$ or $Z_{\rm BRI}>0.8$, there is not such a country with $Z_{EUR}>0.8$. Consequently, the countries with a TCP for EUR are moderately bounded to it and a non negligible part of them are on the brin98k of a transition towards the BRI sector. Moreover, in 2019, the countries in the USD and BRI sectors are principally located in the $Z_{\rm EUR}<0.2$ sector indicating that the countries with a TCP for USD or BRI are very weakly influenced by the EUR.

The time evolutions of the USD, EUR and BRI groups, shown in Fig.~\ref{fig11}, share strong similarities with those of the USD, BRI and swing groups for the two currencies model (Fig.~\ref{fig4}). Again, the swing group is replaced by the EUR group.

As a summary, our currency model allows to probe how influential are economical blocks of countries in the frame of international trade. Another example of application is, e.g., how influential are the OPEC+ countries if the price of oil and petroleum products are multiplied by a factor $K$. In this last example, we keep the Anglo-Saxon countries trading in USD, China in CNY and 11 OPEC+ countries (listed in Tab.~\ref{tab1}) in OPE an hypothetical currency. Contrarily to the hypothetical BRICS currency, no political entity claims the creation of such a petrocurrency. Nonetheless, in our model, a country with a TCP for OPE will have significant trade ties with the OPEC+ countries. Once the steady state of the asynchronous Monte Carlo designation of the countries' TCPs is achieved, we obtain 4 groups (see Fig.~\ref{figA8}): the USD group, the CNY group, the OPE group, and a swing group. This latter group contains countries which, depending on the initial distribution of the TCPs, swings between the 3 currencies, namely, USD, CNY and OPE. Besides the fact, already mentioned through the BRICS+ case, that China gained a prominent economical influence during the last decade, we observe that the OPEC+ is not enough influential to constrain extra-OPEC+ countries to always adopt OPE as TCP, i.e., whatever is the initial TCP distribution over the countries. Nonetheless, the influence of the OPEC+ countries operates through the swing countries which possibly can adopt OPE as a TCP. For the year 2010, the number of swing countries significantly increases if one multiplies the price of oil and gas products by $K=4$ (Fig.~\ref{figA8}): India and South-East Asian countries becomes swing countries non longer exclusively preferring CNY as trade currency. These increase of the number of swing countries with the factor $K=4$ fades in 2019. From Fig.~\ref{figA9}, we still observe during the last decade the significant growing of the ability of China economy to gather in the CNY \textit{club} many countries trading a non negligible part of the global trade volume, whereas the USD group captures a moderate number of extra-ANGL countries but with a modest trade volume, and the OPE group is unable to capture extra-OPEC+ countries.
Besides the mechanic increase of the OPEC+ traded volume, the multiplication of the oil and gas products prices by a factor $K=4$ induces, from 2010 to 2019, a larger number of countries belonging to the swing group in comparison with the $K=1$ case. As the OPE and USD groups are similar for the $K=1$ and $K=4$ cases, the CNY group capture less efficiently countries in the case of expensive oil and gas case. In 2020, the situation is different as the size of the CNY group is the same for the two cases, being insensitive to the increase of the oil and gas products prices.

\section{Conclusion}

In this work, we extend and generalize the approach of opinion formation
to the WTN which represents an example of
complex directed network. Our method is based on asynchronous Monte Carlo iterations
of local spins, which orientation describes trade preference of a given country to perform trade with one or another currency.
We consider two or three core groups of countries
that always keep trading with a certain currency, e.g. USD, BRI or EUR,
and then we follow the Monte Carlo evolution for spin orientations
which converges to a final steady state orientation of spins
(opinions or trade currency preferences of the countries). The local orientation of spins/opinions, namely the trade currency preference of a country $c$, is determined
by the value of a TCP-score which is a linear combination of terms encoding, both, the relative importance of the volume exchanged between the country $c$ and its economical partners, and the relative importance of the country $c$ economic partners.
We note that a somewhat similar
Monte Carlo iterations process with a certain linear condition for spin orientations
has also been considered in problems of associative memory (see e.g. \cite{hopfield82,benedetti23}).

We mainly consider the case of the currency battle
between 5 Anglo-Saxon countries, with a fixed TCP for USD,
and the 11 BRICS+ countries with a fixed TCP for the hypothetical BRI currency. The obtained results, based on the WTN constructed from the UN Comtrade database \cite{comtrade}
for years 2010 - 2020, show the existence of
two fixed groups of countries with firm TCP for USD and BRI
(these groups do not depend on the initial fractions of the countries with initial TCP for USD or BRI)
and a group of swing countries which may change their TCP depending on the initial fraction of countries with TCP for USD or BRI.
We show that from the year 2016 the BRI group contains more than 50\%
of the world countries. Thus in the year 2020, the BRI group
contains 58\% of all the world countries, the USD group 16\% and
the swing group 26\%.

We also describe the reduced Google matrix analysis of the WTN showing that it establishes integrated trade relations between Anglo-Saxon and BRICS+ countries.

Finally, we extend the model to 3 currencies.
On one hand, we consider the currencies USD, BRI, and EUR which is supported by 9 EU countries. In this case, the TCP distribution, for a given year, converges toward an unique solution which is independent of the initial random world distribution of the TCPs. By comparison with the two currencies model, roughly speaking, most of the swing countries adopt the EUR as trade currency whereas the USD and the BRI groups behave similarly within the two models. On the other hand, in order to analyze the economic influence of the OPEC+ countries, we introduce a fictitious petrocurrency, OPE, which compete against USD and CNY.

As far as we know, the use of opinion formation model to probe the economic influence of currencies and/or economic organizations within international trade is novel. Our model is based solely on the structure of the WTN, namely, the bilateral trade flows between world countries, and does not take account of any political or geopolitical aspects. As a conclusion, we state that nowadays the structure of the international trade would clearly favor the emergence of the hypothetical BRI currency pegged to the BRICS+ economies.
\newpage
\section*{Funding}
This research has been partially supported by the grant NANOX N$^\circ$ ANR-17-EURE-0009 (project MTDINA) in the frame  of the Programme des Investissements d'Avenir, France.  This research has also been supported by the Programme Investissements d’Avenir ANR-15-IDEX-0003.
This work has been supported by the EIPHI Graduate
School (contract ANR-17-EURE-0002) and Bourgogne-Franche-Comté Region.
\section*{Data Availability Statement}
The data presented in this study are available on request from the corresponding author. The world trade data are available at https://comtrade.un.org (accessed on 30 December 2023).
\section*{Acknowledgments}
We thank the UN Statistics Division to grant us a friendly access to the UN Comtrade database.

%%%%%%%%%%%%%%%%%%%%%%%%%%%%%%%%%%%%%%%%%%
%% Optional

%% Only for journal Encyclopedia
%\entrylink{The Link to this entry published on the encyclopedia platform.}
\section*{Abbreviations}
The following abbreviations are used in this manuscript:\\

%\noindent
\begin{tabular}{lp{6.2cm}}
ANGL & Core group of Anglo-Saxon countries\\
BRI & Hypothetical currency pegged to the BRICS economies\\
BRICS & Brazil, Russia, India, China, South Africa\\
BRICS+ & Extended BRICS\\
CNY & Chinese yuan\\
EU & European Union\\
EUR & Euro\\
NATO & North Atlantic Treaty Organization\\
OPE & Hypothetical currency pegged to the major oil and gas country producers\\
OPEC & Organization of the Petroleum Exporting Countries\\
OPEC+ & Extended OPEC\\
REGOMAX & Reduced Google matrix\\
TCP & Trade currency preference\\
UK & United Kingdom\\
UN & United Nations\\
USD & US Dollar\\
WTN & World Trade Network \\
WWW & World Wide Web
\end{tabular}
\renewcommand\thefigure{A\arabic{figure}}  
\setcounter{figure}{0}	
\section*{Appendix A}
\begin{figure}[H]
	\begin{center}
		\includegraphics[width=\columnwidth]{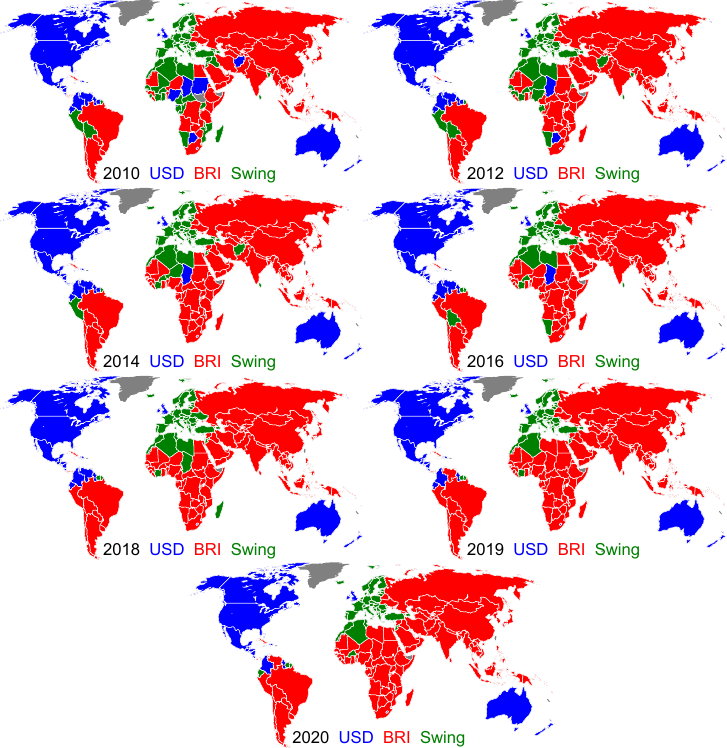}
	\end{center}
	\vglue -0.3cm
	\caption{\label{figA1}World distribution of the trade currency preferences for the years 2010, 2012, 2014, 2016, 2018, 2019 and 2020.
		The countries belonging to the USD group and the BRI group are colored in blue and red, respectively. Those belonging to the swing group are colored in green.
		Countries colored in grey have no trade data reported in the UN Comtrade database for the considered year \cite{comtrade}.
	}
\end{figure}

\begin{figure}[H]
	\begin{center}
		\includegraphics[width=\columnwidth]{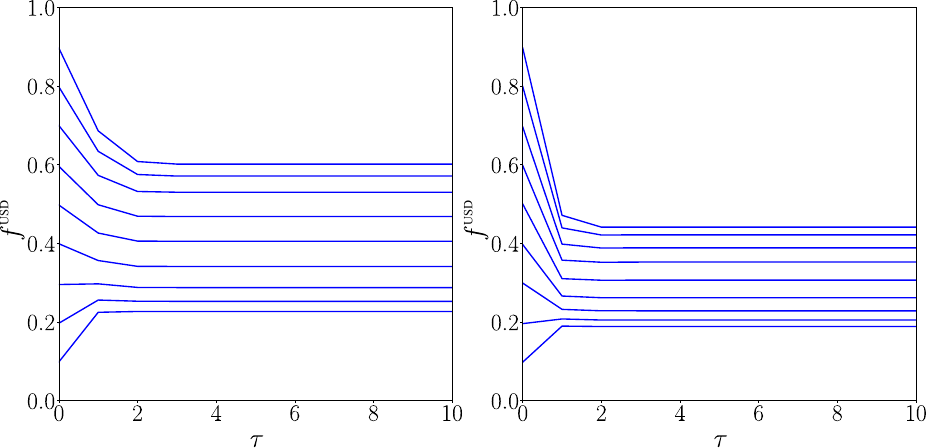}
	\end{center}
	\vglue -0.3cm
	\caption{\label{figA2}Evolution of the fraction of countries preferring USD, $f^{\rm USD}$, with the number of step $\tau$ of the asynchronous Monte Carlo procedure (section~\ref{sec:asynchron}). The fraction $f^{\rm USD}$ is averaged over $10^{4}$ random initial configurations. The left (right) panel concerns the $2010$ WTN ($2019$ WTN).
	}
\end{figure}

\begin{figure}[H]
	\begin{center}
		\includegraphics[width=\columnwidth]{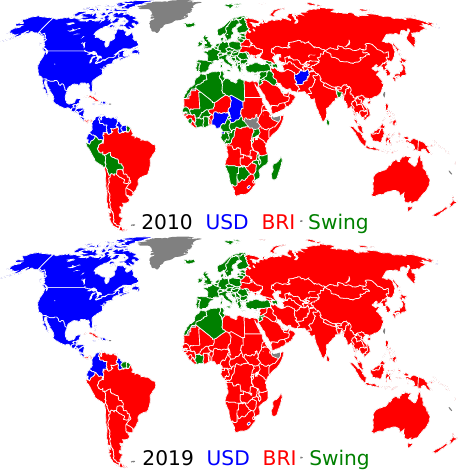}
	\end{center}
	\vglue -0.3cm
	\caption{\label{figA3}Same as Fig.~\ref{fig1} but with the ANGL group only containing the USA. 
		%	World distribution of the trade currency preferences for the years 2010 (top) and 2019 (bottom).
		%	%  and $G_{1}$-option of core groups of Table~\ref{tab1}.
		%	The countries belonging to the USD group and the BRI group are colored in blue and red, respectively. Those belonging to the swing group are colored in green.
		%	Countries colored in grey have no trade data reported in the UN Comtrade database for the considered year \cite{comtrade}.
		%%	The world distribution of trade currency preferences for 2012, 2014, 2016, 2018 and 2020 are presented in Fig.~\ref{figA1}.
	}
\end{figure}

\begin{figure}[H]
	\begin{center}
		\includegraphics[width=\columnwidth]{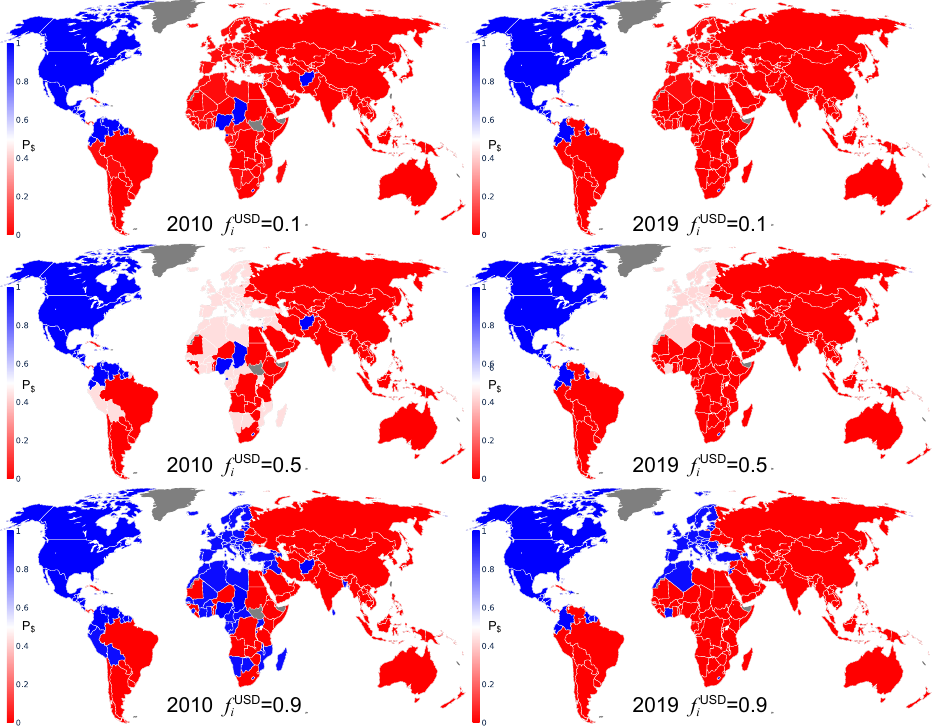}
	\end{center}
	\vglue -0.3cm
	\caption{\label{figA4}Same as Fig.~\ref{fig2} but with the ANGL group only containing the USA.
		%World distribution of the probability $P_\$$ that a country choose USD as its trade currency for 2010 (left) and 2019 (right), and for $f_{i}^{\rm USD}=0.1$ (top), $0.5$ (center) and $0.9$ (bottom). The colors range from red for countries which always have a TCP for BRI ($P_\$=0$) to blue for countries which always have a TCP for USD ($P_\$=1$).
		%Countries colored in grey have no trade data reported in the UN Comtrade database \cite{comtrade}.
	}
\end{figure}

\begin{figure}[H]
	\begin{center}
		\includegraphics[width=\columnwidth]{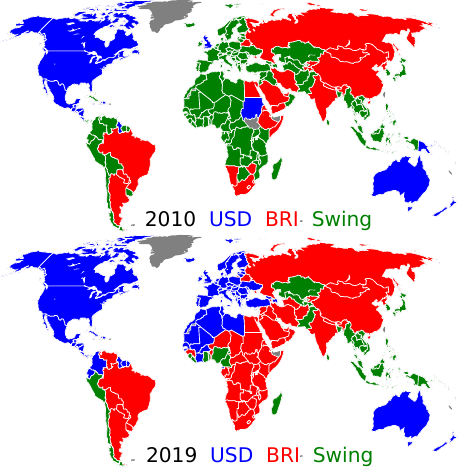}
	\end{center}
	\vglue -0.3cm
	\caption{\label{figA5}Same as Fig.~\ref{fig1} but for $\left(P_{c'}+P^*_{c'}\right)=1$ in the TCP-score (\ref{eq1}).
		World distribution of the trade currency preferences for the years 2010 (top) and 2019 (bottom) taking $f_i^{\rm USD}=0.9$ as the initial fraction of countries preferring USD.
		%  and $G_{1}$-option of core groups of Table~\ref{tab1}.
		The countries belonging to the USD group and the BRI group are colored in blue and red, respectively. Those belonging to the swing group are colored in green.
		Countries colored in grey have no trade data reported in the UN Comtrade database for the considered year \cite{comtrade}.}
\end{figure}

\begin{figure}[H]
	\begin{center}
		\includegraphics[width=\columnwidth]{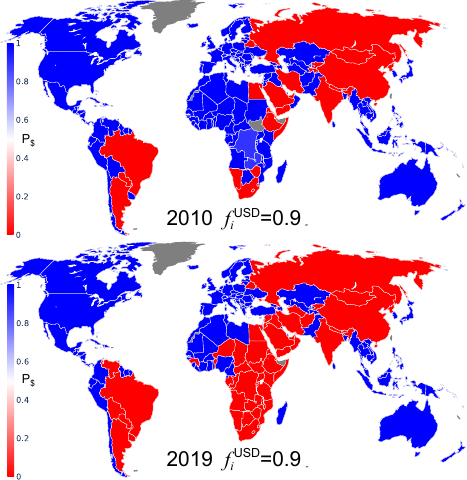}
	\end{center}
	\vglue -0.3cm
	\caption{\label{figA6}Same as Fig.~\ref{fig2} but for $\left(P_{c'}+P^*_{c'}\right)=1$ in the TCP-score (\ref{eq1}). World distribution of the probability $P_\$$ that a country choose USD as its trade currency for 2010 (top) and 2019 (bottom), and for $f_{i}^{\rm USD}=0.9$. The colors range from red for countries which always have a TCP for BRI ($P_\$=0$) to blue for countries which always have a TCP for USD ($P_\$=1$).
		Countries colored in grey have no trade data reported in the UN Comtrade database for the considered year \cite{comtrade}.}
\end{figure}

\begin{figure}[H]
	\begin{center}
		\includegraphics[width=\columnwidth]{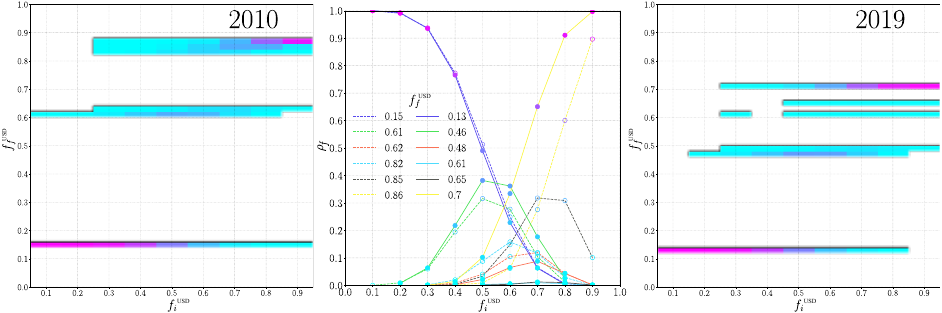}
	\end{center}
	\vglue -0.3cm
	\caption{\label{figA7}Same as Fig.~\ref{fig3} but for $\left(P_{c'}+P^*_{c'}\right)=1$ in the TCP-score (\ref{eq1}).
		Final fraction $f_{f}^{\rm USD}$ of countries with a trade currency preference for USD versus the initial fraction $f_{i}^{\rm USD}$ of these countries for years 2010 (left panel) and 2019 (right panel). The number of final states and their corresponding value $f_f^{\rm USD}$ are initial fraction $f_i^{\rm USD}$ dependent. The color of the points represents the ratio of Monte Carlo process with the corresponding final state $\rho_{f_{f}^{\rm USD}}(f_{i}^{\rm USD})$, low ratio in cold blue and high ration in violet.
		The central panel shows evolution of $\rho_{f_{f}^{\rm USD}}$ with $f_{i}^{\rm USD}$.
		Each line corresponds to a given $f_f^{\rm USD}$. Full (empty) circles and solid (dashed) lines correspond to the year 2019 (2010).}
\end{figure}

\begin{figure}[H]
	\begin{center}
		\includegraphics[width=\columnwidth]{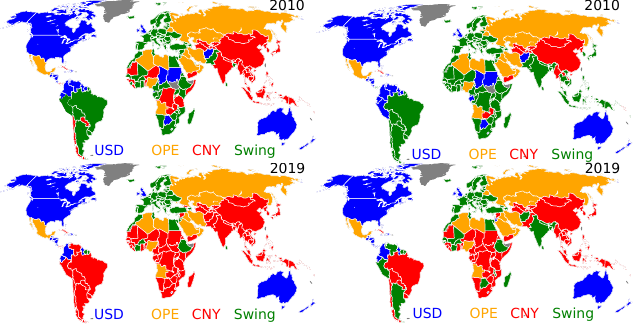}
	\end{center}
	\vglue -0.3cm
	\caption{\label{figA8}World distribution of the trade currency preferences for the years 2010 (top) and 2019 (bottom). The price of oil and gas products is multiplied by $K=1$ (left) and $K=4$ (right).
		%  and $G_{1}$-option of core groups of Table~\ref{tab1}.
		The countries belonging to the USD group, the CNY group, and the OPE group are colored in blue, red, and gold, respectively. Those belonging to the swing group are colored in green.
		Countries colored in grey have no trade data reported in the UN Comtrade database for the considered year \cite{comtrade}.
		%		The world distribution of trade currency preferences for 2012, 2014, 2016, 2018 and 2020 are presented in Fig.~\ref{figA1}.
	}
\end{figure}

\begin{figure}[H]
	\begin{center}
		\includegraphics[width=\columnwidth]{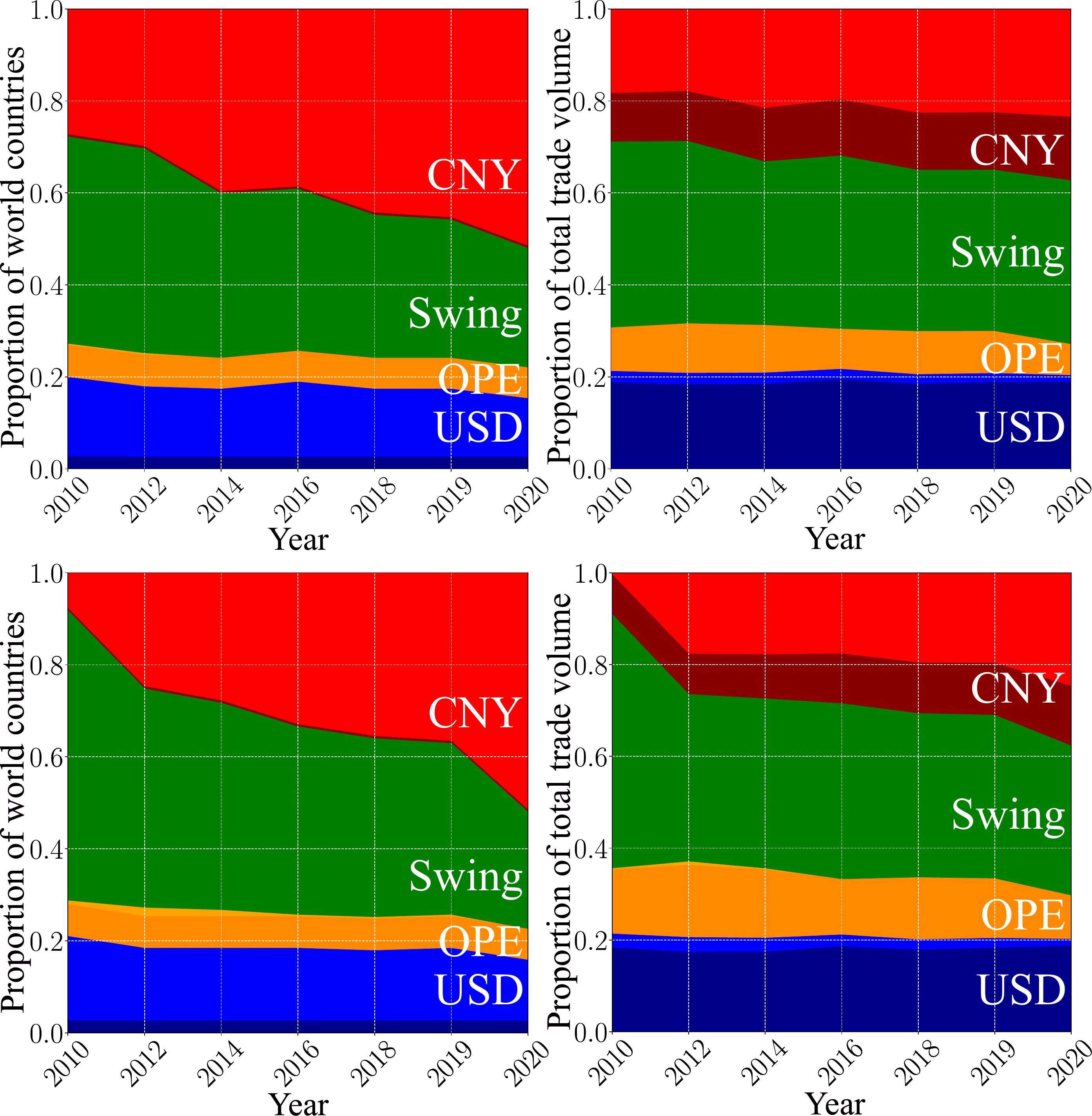}
	\end{center}
	\vglue -0.3cm
	\caption{\label{figA9}Time evolution of the size of the trade currency preference groups.
		The width of a given band corresponds to the corresponding fraction of world countries in a TCP group (left panel) and to the corresponding fraction of the total trade volume generated by this group (right panel).
		The price of oil and gas products is multiplied by $K=1$ (top) and $K=4$ (bottom).
		The USD group is colored in blue, the CNY group in red, the OPE group in gold, and the swing group in green. Within the USD (CNY) [OPE] group, the proportion corresponding to the Anglo-Saxon countries (China) [OPEC+ countries] is shown in dark blue (dark red) [dark gold].}
\end{figure}
\clearpage

\section*{References}
\bibliography{brics2024}

%apsrev4-2.bst 2019-01-14 (MD) hand-edited version of apsrev4-1.bst
%Control: key (0)
%Control: author (8) initials jnrlst
%Control: editor formatted (1) identically to author
%Control: production of article title (0) allowed
%Control: page (0) single
%Control: year (1) truncated
%Control: production of eprint (0) enabled
\begin{thebibliography}{40}%
\makeatletter
\providecommand \@ifxundefined [1]{%
 \@ifx{#1\undefined}
}%
\providecommand \@ifnum [1]{%
 \ifnum #1\expandafter \@firstoftwo
 \else \expandafter \@secondoftwo
 \fi
}%
\providecommand \@ifx [1]{%
 \ifx #1\expandafter \@firstoftwo
 \else \expandafter \@secondoftwo
 \fi
}%
\providecommand \natexlab [1]{#1}%
\providecommand \enquote  [1]{``#1''}%
\providecommand \bibnamefont  [1]{#1}%
\providecommand \bibfnamefont [1]{#1}%
\providecommand \citenamefont [1]{#1}%
\providecommand \href@noop [0]{\@secondoftwo}%
\providecommand \href [0]{\begingroup \@sanitize@url \@href}%
\providecommand \@href[1]{\@@startlink{#1}\@@href}%
\providecommand \@@href[1]{\endgroup#1\@@endlink}%
\providecommand \@sanitize@url [0]{\catcode `\\12\catcode `\$12\catcode
  `\&12\catcode `\#12\catcode `\^12\catcode `\_12\catcode `\%12\relax}%
\providecommand \@@startlink[1]{}%
\providecommand \@@endlink[0]{}%
\providecommand \url  [0]{\begingroup\@sanitize@url \@url }%
\providecommand \@url [1]{\endgroup\@href {#1}{\urlprefix }}%
\providecommand \urlprefix  [0]{URL }%
\providecommand \Eprint [0]{\href }%
\providecommand \doibase [0]{https://doi.org/}%
\providecommand \selectlanguage [0]{\@gobble}%
\providecommand \bibinfo  [0]{\@secondoftwo}%
\providecommand \bibfield  [0]{\@secondoftwo}%
\providecommand \translation [1]{[#1]}%
\providecommand \BibitemOpen [0]{}%
\providecommand \bibitemStop [0]{}%
\providecommand \bibitemNoStop [0]{.\EOS\space}%
\providecommand \EOS [0]{\spacefactor3000\relax}%
\providecommand \BibitemShut  [1]{\csname bibitem#1\endcsname}%
\let\auto@bib@innerbib\@empty
%</preamble>
\bibitem [{\citenamefont {Zaller}(1992)}]{zaller92}%
  \BibitemOpen
  \bibfield  {author} {\bibinfo {author} {\bibfnamefont {J.~R.}\ \bibnamefont
  {Zaller}},\ }\href@noop {} {\emph {\bibinfo {title} {The Nature and Origins
  of Mass Opinion}}},\ Cambridge Studies in Public Opinion and Political
  Psychology\ (\bibinfo  {publisher} {Cambridge University Press},\ \bibinfo
  {year} {1992})\BibitemShut {NoStop}%
\bibitem [{\citenamefont {Castellano}\ \emph {et~al.}(2009)\citenamefont
  {Castellano}, \citenamefont {Fortunato},\ and\ \citenamefont
  {Loreto}}]{fortunato09}%
  \BibitemOpen
  \bibfield  {author} {\bibinfo {author} {\bibfnamefont {C.}~\bibnamefont
  {Castellano}}, \bibinfo {author} {\bibfnamefont {S.}~\bibnamefont
  {Fortunato}},\ and\ \bibinfo {author} {\bibfnamefont {V.}~\bibnamefont
  {Loreto}},\ }\bibfield  {title} {\bibinfo {title} {Statistical physics of
  social dynamics},\ }\href {https://doi.org/10.1103/RevModPhys.81.591}
  {\bibfield  {journal} {\bibinfo  {journal} {Rev. Mod. Phys.}\ }\textbf
  {\bibinfo {volume} {81}},\ \bibinfo {pages} {591} (\bibinfo {year}
  {2009})}\BibitemShut {NoStop}%
\bibitem [{\citenamefont {Dorogovtsev}(2010)}]{dorogovtsev10}%
  \BibitemOpen
  \bibfield  {author} {\bibinfo {author} {\bibfnamefont {S.}~\bibnamefont
  {Dorogovtsev}},\ }\href@noop {} {\emph {\bibinfo {title} {Lectures in Complex
  Networks}}}\ (\bibinfo  {publisher} {Oxford University Press},\ \bibinfo
  {year} {2010})\BibitemShut {NoStop}%
\bibitem [{\citenamefont {Galam}(1986)}]{galam86}%
  \BibitemOpen
  \bibfield  {author} {\bibinfo {author} {\bibfnamefont {S.}~\bibnamefont
  {Galam}},\ }\bibfield  {title} {\bibinfo {title} {Majority rule, hierarchical
  structures, and democratic totalitarianism: A statistical approach},\ }\href
  {https://doi.org/https://doi.org/10.1016/0022-2496(86)90019-2} {\bibfield
  {journal} {\bibinfo  {journal} {Journal of Mathematical Psychology}\ }\textbf
  {\bibinfo {volume} {30}},\ \bibinfo {pages} {426} (\bibinfo {year}
  {1986})}\BibitemShut {NoStop}%
\bibitem [{\citenamefont {Liggett}(1999)}]{liggett99}%
  \BibitemOpen
  \bibfield  {author} {\bibinfo {author} {\bibfnamefont {T.}~\bibnamefont
  {Liggett}},\ }\href {https://doi.org/10.1007/978-3-662-03990-8} {\emph
  {\bibinfo {title} {Stochastic Interacting Systems: Contact, Voter and
  Exclusion Processes}}},\ Grundlehren der mathematischen Wissenschaften\
  (\bibinfo  {publisher} {Springer Berlin Heidelberg},\ \bibinfo {address}
  {Berlin},\ \bibinfo {year} {1999})\BibitemShut {NoStop}%
\bibitem [{\citenamefont {Sznajd-Weron}\ and\ \citenamefont
  {Sznajd}(2000)}]{sznajd00}%
  \BibitemOpen
  \bibfield  {author} {\bibinfo {author} {\bibfnamefont {K.}~\bibnamefont
  {Sznajd-Weron}}\ and\ \bibinfo {author} {\bibfnamefont {J.}~\bibnamefont
  {Sznajd}},\ }\bibfield  {title} {\bibinfo {title} {Opinion evolution in
  closed community},\ }\href {https://doi.org/10.1142/S0129183100000936}
  {\bibfield  {journal} {\bibinfo  {journal} {International Journal of Modern
  Physics C}\ }\textbf {\bibinfo {volume} {11}},\ \bibinfo {pages} {1157}
  (\bibinfo {year} {2000})},\ \Eprint
  {https://arxiv.org/abs/https://doi.org/10.1142/S0129183100000936}
  {https://doi.org/10.1142/S0129183100000936} \BibitemShut {NoStop}%
\bibitem [{\citenamefont {Galam}(2005)}]{galam05}%
  \BibitemOpen
  \bibfield  {author} {\bibinfo {author} {\bibfnamefont {S.}~\bibnamefont
  {Galam}},\ }\bibfield  {title} {\bibinfo {title} {Local dynamics vs. social
  mechanisms: A unifying frame},\ }\href
  {https://doi.org/10.1209/epl/i2004-10526-5} {\bibfield  {journal} {\bibinfo
  {journal} {Europhys. Lett.}\ }\textbf {\bibinfo {volume} {70}},\ \bibinfo
  {pages} {705} (\bibinfo {year} {2005})}\BibitemShut {NoStop}%
\bibitem [{\citenamefont {Sood}\ and\ \citenamefont {Redner}(2005)}]{sood05}%
  \BibitemOpen
  \bibfield  {author} {\bibinfo {author} {\bibfnamefont {V.}~\bibnamefont
  {Sood}}\ and\ \bibinfo {author} {\bibfnamefont {S.}~\bibnamefont {Redner}},\
  }\bibfield  {title} {\bibinfo {title} {Voter model on heterogeneous graphs},\
  }\href {https://doi.org/10.1103/PhysRevLett.94.178701} {\bibfield  {journal}
  {\bibinfo  {journal} {Phys. Rev. Lett.}\ }\textbf {\bibinfo {volume} {94}},\
  \bibinfo {pages} {178701} (\bibinfo {year} {2005})}\BibitemShut {NoStop}%
\bibitem [{\citenamefont {Watts}\ and\ \citenamefont {Dodds}(2007)}]{watts07}%
  \BibitemOpen
  \bibfield  {author} {\bibinfo {author} {\bibfnamefont {D.~J.}\ \bibnamefont
  {Watts}}\ and\ \bibinfo {author} {\bibfnamefont {P.~S.}\ \bibnamefont
  {Dodds}},\ }\bibfield  {title} {\bibinfo {title} {{Influentials, Networks,
  and Public Opinion Formation}},\ }\href {https://doi.org/10.1086/518527}
  {\bibfield  {journal} {\bibinfo  {journal} {Journal of Consumer Research}\
  }\textbf {\bibinfo {volume} {34}},\ \bibinfo {pages} {441} (\bibinfo {year}
  {2007})},\ \Eprint {https://arxiv.org/abs/https://doi.org/10.1086/518527}
  {https://doi.org/10.1086/518527} \BibitemShut {NoStop}%
\bibitem [{\citenamefont {Galam}(2008)}]{galam08}%
  \BibitemOpen
  \bibfield  {author} {\bibinfo {author} {\bibfnamefont {S.}~\bibnamefont
  {Galam}},\ }\bibfield  {title} {\bibinfo {title} {Sociophysics: a review of
  galam models},\ }\href {https://doi.org/10.1142/S0129183108012297} {\bibfield
   {journal} {\bibinfo  {journal} {International Journal of Modern Physics C}\
  }\textbf {\bibinfo {volume} {19}},\ \bibinfo {pages} {409} (\bibinfo {year}
  {2008})},\ \Eprint
  {https://arxiv.org/abs/https://doi.org/10.1142/S0129183108012297}
  {https://doi.org/10.1142/S0129183108012297} \BibitemShut {NoStop}%
\bibitem [{\citenamefont {Schmittmann}\ and\ \citenamefont
  {Mukhopadhyay}(2010)}]{schmittmann10}%
  \BibitemOpen
  \bibfield  {author} {\bibinfo {author} {\bibfnamefont {B.}~\bibnamefont
  {Schmittmann}}\ and\ \bibinfo {author} {\bibfnamefont {A.}~\bibnamefont
  {Mukhopadhyay}},\ }\bibfield  {title} {\bibinfo {title} {Opinion formation on
  adaptive networks with intensive average degree},\ }\href
  {https://doi.org/10.1103/PhysRevE.82.066104} {\bibfield  {journal} {\bibinfo
  {journal} {Phys. Rev. E}\ }\textbf {\bibinfo {volume} {82}},\ \bibinfo
  {pages} {066104} (\bibinfo {year} {2010})}\BibitemShut {NoStop}%
\bibitem [{\citenamefont {Biswas}\ \emph {et~al.}(2012)\citenamefont {Biswas},
  \citenamefont {Chatterjee},\ and\ \citenamefont {Sen}}]{biswas12}%
  \BibitemOpen
  \bibfield  {author} {\bibinfo {author} {\bibfnamefont {S.}~\bibnamefont
  {Biswas}}, \bibinfo {author} {\bibfnamefont {A.}~\bibnamefont {Chatterjee}},\
  and\ \bibinfo {author} {\bibfnamefont {P.}~\bibnamefont {Sen}},\ }\bibfield
  {title} {\bibinfo {title} {Disorder induced phase transition in kinetic
  models of opinion dynamics},\ }\href
  {https://doi.org/https://doi.org/10.1016/j.physa.2012.01.046} {\bibfield
  {journal} {\bibinfo  {journal} {Phys. A Stat. Mech. Its Appl.}\ }\textbf
  {\bibinfo {volume} {391}},\ \bibinfo {pages} {3257} (\bibinfo {year}
  {2012})}\BibitemShut {NoStop}%
\bibitem [{\citenamefont {Crokidakis}(2014)}]{crokidakis14}%
  \BibitemOpen
  \bibfield  {author} {\bibinfo {author} {\bibfnamefont {N.}~\bibnamefont
  {Crokidakis}},\ }\bibfield  {title} {\bibinfo {title} {Phase transition in
  kinetic exchange opinion models with independence},\ }\href
  {https://doi.org/https://doi.org/10.1016/j.physleta.2014.04.028} {\bibfield
  {journal} {\bibinfo  {journal} {Phys. Lett. A}\ }\textbf {\bibinfo {volume}
  {378}},\ \bibinfo {pages} {1683} (\bibinfo {year} {2014})}\BibitemShut
  {NoStop}%
\bibitem [{\citenamefont {Biswas}\ \emph {et~al.}(2023)\citenamefont {Biswas},
  \citenamefont {Chatterjee}, \citenamefont {Sen}, \citenamefont {Mukherjee},\
  and\ \citenamefont {Chakrabarti}}]{biswas23}%
  \BibitemOpen
  \bibfield  {author} {\bibinfo {author} {\bibfnamefont {S.}~\bibnamefont
  {Biswas}}, \bibinfo {author} {\bibfnamefont {A.}~\bibnamefont {Chatterjee}},
  \bibinfo {author} {\bibfnamefont {P.}~\bibnamefont {Sen}}, \bibinfo {author}
  {\bibfnamefont {S.}~\bibnamefont {Mukherjee}},\ and\ \bibinfo {author}
  {\bibfnamefont {B.~K.}\ \bibnamefont {Chakrabarti}},\ }\bibfield  {title}
  {\bibinfo {title} {Social dynamics through kinetic exchange: The {BChS}
  model},\ }\href {https://doi.org/https://doi.org/10.3389/fphy.2023.1196745}
  {\bibfield  {journal} {\bibinfo  {journal} {Front. Phys.}\ }\textbf {\bibinfo
  {volume} {11}} (\bibinfo {year} {2023})}\BibitemShut {NoStop}%
\bibitem [{\citenamefont {Kandiah}\ and\ \citenamefont
  {Shepelyansky}(2012)}]{kandiah12}%
  \BibitemOpen
  \bibfield  {author} {\bibinfo {author} {\bibfnamefont {V.}~\bibnamefont
  {Kandiah}}\ and\ \bibinfo {author} {\bibfnamefont {D.}~\bibnamefont
  {Shepelyansky}},\ }\bibfield  {title} {\bibinfo {title} {{PageRank model of
  opinion formation on social networks}},\ }\href
  {https://doi.org/10.1016/j.physa.2012.06.047} {\bibfield  {journal} {\bibinfo
   {journal} {{Physica A}}\ }\textbf {\bibinfo {volume} {391}},\ \bibinfo
  {pages} {5779} (\bibinfo {year} {2012})}\BibitemShut {NoStop}%
\bibitem [{\citenamefont {Eom}\ and\ \citenamefont
  {Shepelyansky}(2015)}]{eom15}%
  \BibitemOpen
  \bibfield  {author} {\bibinfo {author} {\bibfnamefont {Y.-H.}\ \bibnamefont
  {Eom}}\ and\ \bibinfo {author} {\bibfnamefont {D.~L.}\ \bibnamefont
  {Shepelyansky}},\ }\bibfield  {title} {\bibinfo {title} {Opinion formation
  driven by pagerank node influence on directed networks},\ }\href
  {https://doi.org/https://doi.org/10.1016/j.physa.2015.05.095} {\bibfield
  {journal} {\bibinfo  {journal} {Physica A: Statistical Mechanics and its
  Applications}\ }\textbf {\bibinfo {volume} {436}},\ \bibinfo {pages} {707}
  (\bibinfo {year} {2015})}\BibitemShut {NoStop}%
\bibitem [{\citenamefont {Brin}\ and\ \citenamefont {Page}(1998)}]{brin98}%
  \BibitemOpen
  \bibfield  {author} {\bibinfo {author} {\bibfnamefont {S.}~\bibnamefont
  {Brin}}\ and\ \bibinfo {author} {\bibfnamefont {L.}~\bibnamefont {Page}},\
  }\bibfield  {title} {\bibinfo {title} {{The anatomy of a large-scale
  hypertextual Web search engine}},\ }\href@noop {} {\bibfield  {journal}
  {\bibinfo  {journal} {{Computer Networks and ISDN Systems}}\ }\textbf
  {\bibinfo {volume} {30}},\ \bibinfo {pages} {107} (\bibinfo {year}
  {1998})}\BibitemShut {NoStop}%
\bibitem [{\citenamefont {Langville}\ and\ \citenamefont
  {Meyer}(2006)}]{langville06}%
  \BibitemOpen
  \bibfield  {author} {\bibinfo {author} {\bibfnamefont {A.}~\bibnamefont
  {Langville}}\ and\ \bibinfo {author} {\bibfnamefont {C.}~\bibnamefont
  {Meyer}},\ }\href@noop {} {\emph {\bibinfo {title} {{Google's PageRank and
  beyond: the science of search engine rankings}}}}\ (\bibinfo  {publisher}
  {Princeton University Press},\ \bibinfo {address} {Princeton},\ \bibinfo
  {year} {2006})\BibitemShut {NoStop}%
\bibitem [{\citenamefont {Frahm}\ and\ \citenamefont
  {Shepelyansky}(2019)}]{frahm19}%
  \BibitemOpen
  \bibfield  {author} {\bibinfo {author} {\bibfnamefont {K.~M.}\ \bibnamefont
  {Frahm}}\ and\ \bibinfo {author} {\bibfnamefont {D.~L.}\ \bibnamefont
  {Shepelyansky}},\ }\bibfield  {title} {\bibinfo {title} {Ising-pagerank model
  of opinion formation on social networks},\ }\href
  {https://doi.org/https://doi.org/10.1016/j.physa.2019.121069} {\bibfield
  {journal} {\bibinfo  {journal} {Physica A: Statistical Mechanics and its
  Applications}\ }\textbf {\bibinfo {volume} {526}},\ \bibinfo {pages} {121069}
  (\bibinfo {year} {2019})}\BibitemShut {NoStop}%
\bibitem [{\citenamefont {Ermann}\ \emph {et~al.}(2015)\citenamefont {Ermann},
  \citenamefont {Frahm},\ and\ \citenamefont {Shepelyansky}}]{ermann15a}%
  \BibitemOpen
  \bibfield  {author} {\bibinfo {author} {\bibfnamefont {L.}~\bibnamefont
  {Ermann}}, \bibinfo {author} {\bibfnamefont {K.~M.}\ \bibnamefont {Frahm}},\
  and\ \bibinfo {author} {\bibfnamefont {D.~L.}\ \bibnamefont {Shepelyansky}},\
  }\bibfield  {title} {\bibinfo {title} {Google matrix analysis of directed
  networks},\ }\href {https://doi.org/10.1103/RevModPhys.87.1261} {\bibfield
  {journal} {\bibinfo  {journal} {Rev. Mod. Phys.}\ }\textbf {\bibinfo {volume}
  {87}},\ \bibinfo {pages} {1261} (\bibinfo {year} {2015})}\BibitemShut
  {NoStop}%
\bibitem [{\citenamefont {{United Nations Statistics Division}}()}]{comtrade}%
  \BibitemOpen
  \bibfield  {author} {\bibinfo {author} {\bibnamefont {{United Nations
  Statistics Division}}},\ }\href {http://comtrade.un.org/db/} {\bibinfo
  {title} {United nations commodity trade statistics database}},\ \bibinfo
  {note} {[Online; accessed 30.12.2023]}\BibitemShut {NoStop}%
\bibitem [{\citenamefont {Benedictis}\ and\ \citenamefont
  {Tajoli}(2011)}]{benedictis11}%
  \BibitemOpen
  \bibfield  {author} {\bibinfo {author} {\bibfnamefont {L.~D.}\ \bibnamefont
  {Benedictis}}\ and\ \bibinfo {author} {\bibfnamefont {L.}~\bibnamefont
  {Tajoli}},\ }\bibfield  {title} {\bibinfo {title} {{The World Trade
  Network}},\ }\href {https://doi.org/10.1111/j.1467-9701.2011.01360.x}
  {\bibfield  {journal} {\bibinfo  {journal} {The World Economy}\ }\textbf
  {\bibinfo {volume} {34}},\ \bibinfo {pages} {1417} (\bibinfo {year}
  {2011})}\BibitemShut {NoStop}%
\bibitem [{\citenamefont {Ermann}\ and\ \citenamefont
  {Shepelyansky}(2011)}]{ermann11}%
  \BibitemOpen
  \bibfield  {author} {\bibinfo {author} {\bibfnamefont {L.}~\bibnamefont
  {Ermann}}\ and\ \bibinfo {author} {\bibfnamefont {D.}~\bibnamefont
  {Shepelyansky}},\ }\bibfield  {title} {\bibinfo {title} {{Google matrix of
  the world trade network}},\ }\href
  {https://doi.org/10.12693/APhysPolA.120.A-158} {\bibfield  {journal}
  {\bibinfo  {journal} {{Acta Physica Polonica A}}\ }\textbf {\bibinfo {volume}
  {120}},\ \bibinfo {pages} {A158} (\bibinfo {year} {2011})}\BibitemShut
  {NoStop}%
\bibitem [{\citenamefont {Ermann}\ and\ \citenamefont
  {Shepelyansky}(2015)}]{ermann15b}%
  \BibitemOpen
  \bibfield  {author} {\bibinfo {author} {\bibfnamefont {L.}~\bibnamefont
  {Ermann}}\ and\ \bibinfo {author} {\bibfnamefont {D.}~\bibnamefont
  {Shepelyansky}},\ }\bibfield  {title} {\bibinfo {title} {{Google matrix
  analysis of the multiproduct world trade network}},\ }\href
  {https://doi.org/10.1140/epjb/e2015-60047-0} {\bibfield  {journal} {\bibinfo
  {journal} {{Eur. Phys. J. B}}\ }\textbf {\bibinfo {volume} {88}},\ \bibinfo
  {pages} {84} (\bibinfo {year} {2015})}\BibitemShut {NoStop}%
\bibitem [{\citenamefont {Coquidé}\ \emph {et~al.}(2019)\citenamefont
  {Coquidé}, \citenamefont {Ermann}, \citenamefont {Lages},\ and\
  \citenamefont {Shepelyansky}}]{coquide19}%
  \BibitemOpen
  \bibfield  {author} {\bibinfo {author} {\bibfnamefont {C.}~\bibnamefont
  {Coquidé}}, \bibinfo {author} {\bibfnamefont {L.}~\bibnamefont {Ermann}},
  \bibinfo {author} {\bibfnamefont {J.}~\bibnamefont {Lages}},\ and\ \bibinfo
  {author} {\bibfnamefont {D.}~\bibnamefont {Shepelyansky}},\ }\bibfield
  {title} {\bibinfo {title} {{Influence of petroleum and gas trade on EU
  economies from the reduced Google matrix analysis of UN COMTRADE data}},\
  }\href {https://doi.org/10.1140/epjb/e2019-100132-6} {\bibfield  {journal}
  {\bibinfo  {journal} {{Eur. Phys. J. B}}\ }\textbf {\bibinfo {volume} {92}},\
  \bibinfo {pages} {71} (\bibinfo {year} {2019})}\BibitemShut {NoStop}%
\bibitem [{\citenamefont {Chepelianskii}(2010)}]{chepelianskii10}%
  \BibitemOpen
  \bibfield  {author} {\bibinfo {author} {\bibfnamefont {A.~D.}\ \bibnamefont
  {Chepelianskii}},\ }\href@noop {} {\bibinfo {title} {Towards physical laws
  for software architecture}} (\bibinfo {year} {2010}),\ \Eprint
  {https://arxiv.org/abs/1003.5455} {arXiv:1003.5455 [cs.SE]} \BibitemShut
  {NoStop}%
\bibitem [{\citenamefont {Zhirov}\ \emph {et~al.}(2010)\citenamefont {Zhirov},
  \citenamefont {Zhirov},\ and\ \citenamefont {Shepelyansky}}]{zhirov10}%
  \BibitemOpen
  \bibfield  {author} {\bibinfo {author} {\bibfnamefont {A.}~\bibnamefont
  {Zhirov}}, \bibinfo {author} {\bibfnamefont {O.}~\bibnamefont {Zhirov}},\
  and\ \bibinfo {author} {\bibfnamefont {D.}~\bibnamefont {Shepelyansky}},\
  }\bibfield  {title} {\bibinfo {title} {{Two-dimensional ranking of Wikipedia
  articles}},\ }\href {https://doi.org/10.1140/epjb/e2010-10500-7} {\bibfield
  {journal} {\bibinfo  {journal} {{Eur. Phys. J. B}}\ }\textbf {\bibinfo
  {volume} {77}},\ \bibinfo {pages} {523–531} (\bibinfo {year}
  {2010})}\BibitemShut {NoStop}%
\bibitem [{\citenamefont {Coquidé}\ \emph
  {et~al.}(2023{\natexlab{a}})\citenamefont {Coquidé}, \citenamefont {Lages},\
  and\ \citenamefont {Shepelyansky}}]{coquide23a}%
  \BibitemOpen
  \bibfield  {author} {\bibinfo {author} {\bibfnamefont {C.}~\bibnamefont
  {Coquidé}}, \bibinfo {author} {\bibfnamefont {J.}~\bibnamefont {Lages}},\
  and\ \bibinfo {author} {\bibfnamefont {D.~L.}\ \bibnamefont {Shepelyansky}},\
  }\bibfield  {title} {\bibinfo {title} {Dollar-yuan battle in the world trade
  network},\ }\bibfield  {journal} {\bibinfo  {journal} {Entropy}\ }\textbf
  {\bibinfo {volume} {25}},\ \href {https://doi.org/10.3390/e25020373}
  {10.3390/e25020373} (\bibinfo {year} {2023}{\natexlab{a}})\BibitemShut
  {NoStop}%
\bibitem [{\citenamefont {Hopfield}(2082)}]{hopfield82}%
  \BibitemOpen
  \bibfield  {author} {\bibinfo {author} {\bibfnamefont {J.}~\bibnamefont
  {Hopfield}},\ }\bibfield  {title} {\bibinfo {title} {{Neural networks and
  physical systems with emergent collective computational abilities}},\ }\href
  {https://doi.org/10.1073/pnas.79.8.2554} {\bibfield  {journal} {\bibinfo
  {journal} {{PNAS}}\ }\textbf {\bibinfo {volume} {79}},\ \bibinfo {pages}
  {2554} (\bibinfo {year} {2082})}\BibitemShut {NoStop}%
\bibitem [{\citenamefont {Benedetti}\ \emph {et~al.}(2023)\citenamefont
  {Benedetti}, \citenamefont {Carillo}, \citenamefont {Marinari},\ and\
  \citenamefont {Mèzard}}]{benedetti23}%
  \BibitemOpen
  \bibfield  {author} {\bibinfo {author} {\bibfnamefont {M.}~\bibnamefont
  {Benedetti}}, \bibinfo {author} {\bibfnamefont {L.}~\bibnamefont {Carillo}},
  \bibinfo {author} {\bibfnamefont {E.}~\bibnamefont {Marinari}},\ and\
  \bibinfo {author} {\bibfnamefont {M.}~\bibnamefont {Mèzard}},\ }\href@noop
  {} {\bibinfo {title} {Eigenvector dreaming}} (\bibinfo {year} {2023}),\
  \Eprint {https://arxiv.org/abs/2308.13445} {arXiv:2308.13445
  [cond-mat.dis-nn]} \BibitemShut {NoStop}%
\bibitem [{\citenamefont {Coquidé}\ \emph
  {et~al.}(2023{\natexlab{b}})\citenamefont {Coquidé}, \citenamefont {Lages},\
  and\ \citenamefont {Shepelyansky}}]{coquide23b}%
  \BibitemOpen
  \bibfield  {author} {\bibinfo {author} {\bibfnamefont {C.}~\bibnamefont
  {Coquidé}}, \bibinfo {author} {\bibfnamefont {J.}~\bibnamefont {Lages}},\
  and\ \bibinfo {author} {\bibfnamefont {D.~L.}\ \bibnamefont {Shepelyansky}},\
  }\bibfield  {title} {\bibinfo {title} {Prospects of brics currency dominance
  in international trade},\ }\bibfield  {journal} {\bibinfo  {journal} {Appl
  Netw Sci}\ }\textbf {\bibinfo {volume} {8}},\ \href
  {https://doi.org/10.1007/s41109-023-00590-3} {10.1007/s41109-023-00590-3}
  (\bibinfo {year} {2023}{\natexlab{b}})\BibitemShut {NoStop}%
\bibitem [{\citenamefont {Saint-Etienne}(2018)}]{saint18}%
  \BibitemOpen
  \bibfield  {author} {\bibinfo {author} {\bibfnamefont {C.}~\bibnamefont
  {Saint-Etienne}},\ }\href@noop {} {\emph {\bibinfo {title} {Osons l'Europe
  des nations}}}\ (\bibinfo  {publisher} {L'{\'E}ditions de l'Observatoire},\
  \bibinfo {address} {Paris},\ \bibinfo {year} {2018})\BibitemShut {NoStop}%
\bibitem [{\citenamefont {Loye}\ \emph {et~al.}(2021)\citenamefont {Loye},
  \citenamefont {Ermann},\ and\ \citenamefont {Shepelyansky}}]{loye21}%
  \BibitemOpen
  \bibfield  {author} {\bibinfo {author} {\bibfnamefont {J.}~\bibnamefont
  {Loye}}, \bibinfo {author} {\bibfnamefont {L.}~\bibnamefont {Ermann}},\ and\
  \bibinfo {author} {\bibfnamefont {D.~L.}\ \bibnamefont {Shepelyansky}},\
  }\bibfield  {title} {\bibinfo {title} {World impact of kernel european union
  9 countries from google matrix analysis of the world trade network},\
  }\bibfield  {journal} {\bibinfo  {journal} {Applied Network Science}\
  }\textbf {\bibinfo {volume} {6}},\ \href
  {https://doi.org/10.1007/s41109-021-00380-9} {10.1007/s41109-021-00380-9}
  (\bibinfo {year} {2021})\BibitemShut {NoStop}%
\bibitem [{\citenamefont {Borger}(2023)}]{guardian23}%
  \BibitemOpen
  \bibfield  {author} {\bibinfo {author} {\bibfnamefont {J.}~\bibnamefont
  {Borger}},\ }\href
  {https://www.theguardian.com/business/2023/aug/24/five-brics-nations-announce-admission-of-six-new-countries-to-bloc}
  {\bibinfo {title} {{Brics to more than double with admission of six new
  countries}}} (\bibinfo {year} {2023}),\ \bibinfo {note} {{The Guardian},
  [Online 08.24.2023; Accessed 12.30.2023]}\BibitemShut {NoStop}%
\bibitem [{\citenamefont {{Wikipedia
  contributors}}(2023{\natexlab{a}})}]{bricswiki}%
  \BibitemOpen
  \bibfield  {author} {\bibinfo {author} {\bibnamefont {{Wikipedia
  contributors}}},\ }\href@noop {} {\bibinfo {title} {Brics --- {Wikipedia}{,}
  the free encyclopedia}},\ \bibinfo {howpublished}
  {\url{https://en.wikipedia.org/w/index.php?title=BRICS&oldid=1192546195}}
  (\bibinfo {year} {2023}{\natexlab{a}}),\ \bibinfo {note} {[Online; accessed
  30-December-2023]}\BibitemShut {NoStop}%
\bibitem [{\citenamefont {Frahm}\ \emph {et~al.}(2016)\citenamefont {Frahm},
  \citenamefont {Jaffrès-Runser},\ and\ \citenamefont
  {Shepelyansky}}]{frahm16}%
  \BibitemOpen
  \bibfield  {author} {\bibinfo {author} {\bibfnamefont {K.}~\bibnamefont
  {Frahm}}, \bibinfo {author} {\bibfnamefont {K.}~\bibnamefont
  {Jaffrès-Runser}},\ and\ \bibinfo {author} {\bibfnamefont {D.}~\bibnamefont
  {Shepelyansky}},\ }\bibfield  {title} {\bibinfo {title} {{Wikipedia mining of
  hidden links between political leaders}},\ }\href
  {https://doi.org/10.1140/epjb/e2016-70526-3} {\bibfield  {journal} {\bibinfo
  {journal} {{Eur. Phys. J. B}}\ }\textbf {\bibinfo {volume} {89}},\ \bibinfo
  {pages} {269} (\bibinfo {year} {2016})}\BibitemShut {NoStop}%
\bibitem [{\citenamefont {{Wikipedia
  contributors}}(2023{\natexlab{b}})}]{bretton}%
  \BibitemOpen
  \bibfield  {author} {\bibinfo {author} {\bibnamefont {{Wikipedia
  contributors}}},\ }\href@noop {} {\bibinfo {title} {Bretton woods system ---
  {Wikipedia}{,} the free encyclopedia}},\ \bibinfo {howpublished}
  {\url{https://en.wikipedia.org/w/index.php?title=Bretton_Woods_system&oldid=1187855821}}
  (\bibinfo {year} {2023}{\natexlab{b}}),\ \bibinfo {note} {[Online; accessed
  30-December-2023]}\BibitemShut {NoStop}%
\bibitem [{\citenamefont {Said}\ and\ \citenamefont {Kalin}(2022)}]{wallstrj}%
  \BibitemOpen
  \bibfield  {author} {\bibinfo {author} {\bibfnamefont {S.}~\bibnamefont
  {Said}}\ and\ \bibinfo {author} {\bibfnamefont {S.}~\bibnamefont {Kalin}},\
  }\href
  {https://www.wsj.com/articles/saudi-arabia-considers-accepting-yuan-instead-of-dollars-for-chinese-oil-sales-11647351541}
  {\bibinfo {title} {{Saudi Arabia Considers Accepting Yuan Instead of Dollars
  for Chinese Oil Sales}}} (\bibinfo {year} {2022}),\ \bibinfo {note} {{The
  Wall Street Journal}, [Online 03.15.2022; accessed 12.30.2023]}\BibitemShut
  {NoStop}%
\bibitem [{\citenamefont {Betz}()}]{bricurrency}%
  \BibitemOpen
  \bibfield  {author} {\bibinfo {author} {\bibfnamefont {B.}~\bibnamefont
  {Betz}},\ }\href {https://www.foxbusiness.
  com/markets/brazil-china-strike-trade-deal-agreement-ditch-us-dollar}
  {\bibinfo {title} {{Brazil, China strike trade deal agreement to ditch US
  dollar}}},\ \bibinfo {note} {{FOX Business}, [Online 03.22.2023; accessed
  12.31.2023]}\BibitemShut {NoStop}%
\bibitem [{\citenamefont {Kotelnikova}\ \emph {et~al.}(2022)\citenamefont
  {Kotelnikova}, \citenamefont {Frahm}, \citenamefont {Shepelyansky},\ and\
  \citenamefont {Kunduzova}}]{kotelnikova22}%
  \BibitemOpen
  \bibfield  {author} {\bibinfo {author} {\bibfnamefont {E.}~\bibnamefont
  {Kotelnikova}}, \bibinfo {author} {\bibfnamefont {K.~M.}\ \bibnamefont
  {Frahm}}, \bibinfo {author} {\bibfnamefont {D.~L.}\ \bibnamefont
  {Shepelyansky}},\ and\ \bibinfo {author} {\bibfnamefont {O.}~\bibnamefont
  {Kunduzova}},\ }\bibfield  {title} {\bibinfo {title} {Fibrosis
  protein-protein interactions from google matrix analysis of metacore
  network},\ }\bibfield  {journal} {\bibinfo  {journal} {International Journal
  of Molecular Sciences}\ }\textbf {\bibinfo {volume} {23}},\ \href
  {https://doi.org/10.3390/ijms23010067} {10.3390/ijms23010067} (\bibinfo
  {year} {2022})\BibitemShut {NoStop}%
\end{thebibliography}%

\end{document}